\documentclass[fleqn,usenatbib]{mnras}

\usepackage{newtxtext,newtxmath}
\usepackage[T1]{fontenc}
\usepackage{ae,aecompl}

\usepackage{graphicx}	
\usepackage{amsmath}	
\usepackage{amssymb}	
\usepackage{xspace}
\usepackage{booktabs}
\usepackage[flushleft]{threeparttable}
\usepackage{multirow}
\usepackage{color,soul}

\defcitealias{art:hawkins+2017}{H17}
\defcitealias{art:yu+2018}{Y18}
\newcommand{\numax}{\mbox{$\nu_{\rm max}$}\xspace}
\newcommand{\oozp}{\mbox{$\varpi_{\rm zp}$}\xspace}
\newcommand{\muas}{\mbox{$\mu \rm as$}\xspace}
\newcommand{\murc}{\mbox{$\mu_{\rm RC}$}\xspace}
\newcommand{\sigrc}{\mbox{$\sigma_{\rm RC}$}\xspace}
\newcommand{\fdnu}{\mbox{$f_{\Delta\nu}$}\xspace}
\newcommand{\nstars}{5576 \xspace} 
\newcommand{\logten}{\mbox{$\log_{10}$}\xspace}
\newcommand{\dnu}{\mbox{$\Delta \nu$}\xspace}

\newcommand{\teff}{\mbox{$T_{\rm eff}$}\xspace}
\newcommand{\logg}{\mbox{$\log g$}\xspace}
\newcommand{\feh}{\mbox{$\rm{[Fe/H]}$}\xspace}

\newcommand{\kepler}{\emph{Kepler}\xspace}

\newcommand{\gaia}{\emph{Gaia}\xspace}

\newcommand{\Ks}{\mbox{$K$}\xspace}
\newcommand{\new}[1]{#1}
\newcommand{\nnew}[1]{#1}

\newcommand{\up}[1]{#1} 

\title[Testing systematics with the Red Clump]{Testing asteroseismology with \textit{Gaia} DR2: Hierarchical models of the Red Clump}

\author[O. J. Hall et al.]{Oliver J. Hall$^{1,2}$\thanks{E-mail: ojh251@bham.ac.uk (OJH)}, 
Guy R. Davies$^{1,2}$,
Yvonne P. Elsworth$^{1,2}$,
Andrea Miglio$^{1,2}$, 
\newauthor Timothy R. Bedding$^{3,2}$,
Anthony G. A. Brown$^{4}$,
Saniya Khan$^{1,2}$,
Keith Hawkins$^{5}$,
\newauthor Rafael A. Garc\'ia$^{6, 7}$,
William J. Chaplin$^{1, 2}$,
Thomas S. H. North$^{1}$ 
\\
$^{1}$School of Physics and Astronomy, University of Birmingham, Edgbaston, Birmingham, B15 2TT, UK\\
$^{2}$Stellar Astrophysics Centre, Department of Physics and Astronomy, Aarhus University, Ny Munkegade 120, 8000 Aarhus C, Denmark\\
$^{3}$Sydney Institute for Astronomy, School of Physics, University of Sydney 2006, Australia\\
$^{4}$Leiden Observatory, Leiden University, Niels Bohrweg 2, 2333 CA, Leiden, The Netherlands\\
$^{5}$Department of Astrnomy, The University of Texas at Austin, 2515 Speedway Boulevard, Austin, TX 78712, USA\\
$^{6}$IRFU, CEA, Universit\'e Paris-Saclay, F-91191 Gif-sur-Yvette, France\\
$^{7}$AIM, CEA, CNRS, Universit\'e Paris-Saclay, Universit\'e Paris Diderot, Sorbonne Paris Cit\'e, F-91191 Gif-sur-Yvette, France\\
}

\date{Accepted XXX. Received YYY; in original form ZZZ}

\pubyear{2019}

\begin{document}
\label{firstpage}
\pagerange{\pageref{firstpage}--\pageref{lastpage}}
\maketitle

\begin{abstract}
Asteroseismology provides fundamental stellar parameters independent of distance, but subject to systematics under calibration. \gaia DR2 has provided parallaxes for a billion stars, which are offset by a parallax zero-point (\oozp). Red Clump (RC) stars have a narrow spread in luminosity, thus functioning as standard candles to calibrate these systematics.
This work measures how the magnitude and spread of the RC in the \kepler field are affected by changes to temperature and scaling relations for seismology, and changes to the parallax zero-point for \gaia. We use a sample of \nstars RC stars classified through asteroseismology.
We apply hierarchical Bayesian latent variable models, finding the population level properties of the RC with seismology, and use those as priors on \gaia parallaxes to find \oozp. We then find the position of the RC using published values for \oozp.
We find a seismic temperature insensitive spread of the RC of $\sim0.03\, \rm mag$ in the 2MASS $K$ band and a larger and slightly temperature-dependent spread of $\sim0.13\, \rm mag$ in the \gaia $G$ band. This intrinsic dispersion in the $K$ band provides a distance precision of $\sim 1\%$ for RC stars. Using \gaia data alone, we find a mean zero-point of $-41\pm10\, \mu \rm as$. This offset yields RC absolute magnitudes of $-1.634\pm0.018$ in $K$ and $0.546\pm0.016$ in $G$. Obtaining these same values through seismology would require a global temperature shift of $\sim-70\, K$, which is compatible with known systematics in spectroscopy.\\
\end{abstract}



\section{Introduction}
Since the launch of CoRoT \citep{art:baglin+2006} and \textit{Kepler} \citep{art:borucki+2010}, the use of asteroseismology --- the study of stars' internal physics by observing \up{their modes of oscillation} --- has become a crucial tool for testing fundamental stellar properties. The large quantity of long timeseries photometry from these missions \citep{art:chaplin+miglio2013}, and its distance independent nature, have allowed for measures of precise stellar radii and masses for both red giant stars \citep{art:hekker+2011,art:huber+2011a,art:huber+2014,art:mathur+2016,art:pinsonneault+2014, art:pinsonneault+2018,art:yu+2018} and main sequence stars \citep{art:chaplin+2010, art:chaplin+2011,art:chaplin+2014c}, studies of exoplanets and exoplanet hosts \citep{art:christensen-dalsgaard+2010,art:batalha+2011, art:huber+2013, art:huber+2013a,art:chaplin+2013,  art:silvaaguirre+2015}, internal \& external stellar rotation \citep{art:beck+2012, art:deheuvels+2012, art:deheuvels+2014,  art:mosser+2012, art:davies+2015}, ages of stellar populations \citep{art:miglio+2009,art:miglio+2013,art:casagrande+2014a,art:casagrande+2016,art:stello+2015}, and classifications of stellar types \citep{art:bedding+2011,art:mosser+2012a, art:mosser+2015,art:stello+2013,art:vrard+2016,art:elsworth+2017}, among others.

Many of these works rely on the so-called `direct method': the use of seismic scaling relations related to the two fundamental oscillation parameters, \numax, the frequency of maximum power of the oscillation mode envelope, and \dnu, the spacing between two oscillation modes of equal radial degree. These properties are individually proportional to mass, radius and temperature, and when combined and scaled with solar values, can provide measures of stellar mass, radius and surface gravity \citep{art:kjeldsen+bedding1995}. As such, stellar properties obtained through seismology depend on temperature as well as on the seismic parameters. Besides the direct method, results from seismology can also be obtained by comparing global seismic properties with a grid of models, referred to as `grid modelling', and can be expanded to `detailed modelling', which directly fits observed seismic mode frequencies to the grids \citep{art:metcalfe+2012, art:metcalfe+2014,art:silvaaguirre+2013,art:silvaaguirre+2015,art:davies+2016,art:lund+2017}.

The seismic scaling relations have been thoroughly tested through interferometry \citep{art:white+2013}, astrometry \citep{art:huber+2017}, eclipsing binaries \citep{art:gaulme+2016}, and open clusters \citep{art:miglio+2012}. Theoretically motivated corrections to the \dnu and \numax scaling relations have been proposed to depend on \teff, metallicity, and evolutionary state \citep{art:miglio+2012,art:sharma+2016}, and it is known that a small correction for the mean molecular weight could be needed for the \numax scaling relation \citep{art:belkacem+2013, art:viani+2017}. 


When using the direct method, effective temperatures from spectroscopic analysis are often used (e.g. the APOKASC catalogue; \cite{art:pinsonneault+2014,art:pinsonneault+2018}). However depending on the atmospheric models and temperature scales applied in spectroscopic analysis, inferred values for \teff can vary up to $\sim170\ K$ for Core Helium-Burning (CHeB) stars \citep{art:slumstrup+2018}. While \cite{art:bellinger+2018} have recently shown that these systematic uncertainties can be mitigated through the use of grid modelling for main-sequence and sub-giant stars, the question of which temperature scale for spectroscopy obtains the best value for \teff remains open.\\

Seismic observations can be combined with distance dependent observations, such as astrometry, to improve and calibrate results. The second data release (DR2) of the astrometric \gaia mission \citep{art:gaiacollaboration+2018} recently has provided data for a sample of over one billion targets, with uncertainties largely improved from the first data release \citep[DR1, TGAS][]{art:gaiacollaboration+2016}, allowing for a broader range of science and calibrations \citep{art:zinn+2018}. With DR2 \cite{art:lindegren+2018} suggested a mean global parallax zero-point offset of $-29\ \mu\rm{as}$, in the sense that \gaia parallaxes are too small, using a quasar sample, although it should be noted that the offset varies as a function of colours, magnitude and position on the sky. \cite{art:arenou+2018} computed the parallax difference between DR2 and existing catalogues, as well as prior data for individual targets, and found these on average to be the same order of magnitude as the \cite{art:lindegren+2018} zero-point. \cite{art:riess+2018} used Cepheid variables to derive a zero-point offset of $-46\pm13\ \mu\rm{as}$, \cite{art:stassun+torres2018} used Eclipsing Binaries to find a zero-point of $-83\pm33\ \mu\rm{as}$, and \cite{art:zinn+2018} compared parallaxes to seismic radii to identify a colour- and magnitude-dependent offset of $-52.8\pm2.4(stat.)\pm1(syst.)\ \mu\rm{as}$ for red giant branch stars in the \kepler field. Finally, using analysis of individual seismic mode frequencies for 93 dwarf stars, \cite{art:sahlholdt+silvaaguirre2018} reported an offset in estimated stellar radii equal to a parallax offset of $-35 \pm 16\ \mu \rm as$. \up{As the parallax zero-point offset is known to vary with magnitude, colour, and position in the sky, the differences between these values for the zero-point are expected. Understanding how we quantify the offset is crucial if we want to use \gaia to calibrate asteroseismology and other methods.}\\


One method of testing independent sets of measurements is calculating an observable astronomical property. An example of such a property is the luminosity of the `Red Clump' (RC), an overdensity of red giant stars on the HR-diagram, in bands of absolute magnitude. When stars of masses around \new{$0.7 \lesssim M/M_\odot \lesssim 1.9$ (for $[\rm{M/H}] \simeq 0.07$, upper limit subject to change with metallicity)} ignite helium in their cores, they undergo the He-flash. The \nnew{He-burning} core masses are \nnew{very similar} for these stars, and as their luminosity is mainly determined by the core mass, they will all have similar luminosities, creating a clump of stars on the HR-diagram \citep[][and references therein]{art:girardi2016}. \nnew{Further differences in luminosity and temperature are then effects of metallicity and envelope mass,} \new{and thus the Clump has a relatively small spread.} Stars at lower masses and low metallicities form a horizontal branch at a luminosity similar to the RC, \nnew{whereas stars of masses just above the limit for the He-flash lie at a slightly lower luminosity,} forming a Secondary Red Clump \citep[2CL,][]{art:girardi1999}. At even higher masses, the luminosity becomes a function of stellar mass, and these stars form a vertical structure in the HR-diagram during their CHeB phase.

The luminosity of the RC overdensity may be used as a standard candle given constraints on mass and metallicity \citep{art:cannon1970}, and has recently been used to calibrate \gaia DR1 parallaxes \citep{art:davies+2017}. Also using \gaia DR1 parallaxes, \cite{art:hawkins+2017} (hereafter \citetalias{art:hawkins+2017}) found precise measurements for the RC luminosity in various passbands, including the 2MASS \Ks band, which minimised the spread in luminosity due to mass and metallicity \citep{art:salaris+girardi2002}. With \gaia DR2's improved parallax uncertainties and reduced systematic offset, now is a good time to revisit the RC as a calibrator.

In this work we investigate systematics in both asteroseismology and \gaia \textit{simultaneously}, to see how differences in assumptions for one influence inferences of the other. Using a sample of over 5500 \kepler Red Giant stars in the RC for which parallaxes and seismology are available, we measure the position of the RC population in absolute magnitude in the 2MASS \citep{art:skrutskie+2006} \Ks band, and the \gaia $G$ band. We do this using seismology and parallax (with photometry) independently. \new{Since the distribution of RC stars should be the same for this population, independent of method, a (dis)agreement of the measured positions and spreads of the RC using two independent methods sheds light on systematics in both.} For the seismic method, we test the influence of the temperature scale used to obtain the values of \teff fed into seismic scaling relations, as well as the impact of corrections to the \dnu scaling relation. For the \gaia method, we study how changes in the parallax zero-point offset for \gaia DR2 impact the inferred luminosity of the RC.

This paper is laid out as follows: Section \ref{sec:data} discusses how the data were obtained, and the theory used to calculate our observables. Section \ref{sec:method} discusses how we use hierarchical Bayesian modelling to study the RC. We present our results in Section \ref{sec:results} and discuss them in context of similar work in Section \ref{sec:discussion}, and present our conclusions in Section \ref{sec:conclusions}.

\section{Data} \label{sec:data}
Our aim is to find the intrinsic position and spread of the Red Clump in absolute magnitude for various passbands using two approaches: one using a distance-independent luminosity calculated from asteroseismology, and the other using a magnitude inferred from photometry and \gaia DR2 parallaxes. Since the number of stars with asteroseismic data is significantly lower than those with data in \gaia DR2, this limits our sample. 

For our asteroseismic sample, we used the catalogue of 16,094 oscillating \kepler red giants by \cite{art:yu+2018} (hereafter \citetalias{art:yu+2018}), which contains global oscillation parameters \numax and \dnu, as well as broad evolutionary state classifications, effective temperatures \teff and metallicities \feh taken from \cite{art:mathur+2017}.

We \nnew{re-considered the classification of} all stars labelled as CHeB in the \citetalias{art:yu+2018} catalogue using the method presented in \cite{art:elsworth+2017}. This uses the structure of dipole-mode oscillations in the power spectra to classify stars as belonging to the 2CL, the Red Giant Branch (RGB), or the RC. We obtained light curves for 7437 stars labelled as CHeB in \citetalias{art:yu+2018}, \up{from two sources: the so-called KASOC light curves \citep{art:handberg+lund2014}\footnote{Freely distributed at the KASOC webpage (\url{http://kasoc.phys.au.dk})} and the KEPSEISMIC light curves \citep{art:garcia+2011}\footnote{Freely distributed at the MAST website (\url{https://archive.stsci.edu/prepds/kepseismic/})}. The latter have been produced with larger photometric masks to ensure a better stability at low frequencies, and have been gap-filled using in-painting techniques \citep{art:garcia+2014, art:jofre+2015}.}

Of these 7437 stars, we found that 5668 are RC, 737 are 2CL, and no classification could be found for  \nnew{499} stars. Notably, 533 stars were found to be RGB, disagreeing with the classification listed in \citetalias{art:yu+2018}. \new{This should be discussed in future work, but for the sake of \up{internal consistency of our classifications} we have \nnew{chosen to adopt} \cite{art:elsworth+2017} classification in this work.}

\begin{figure}
\centering
\includegraphics[width=.49\textwidth]{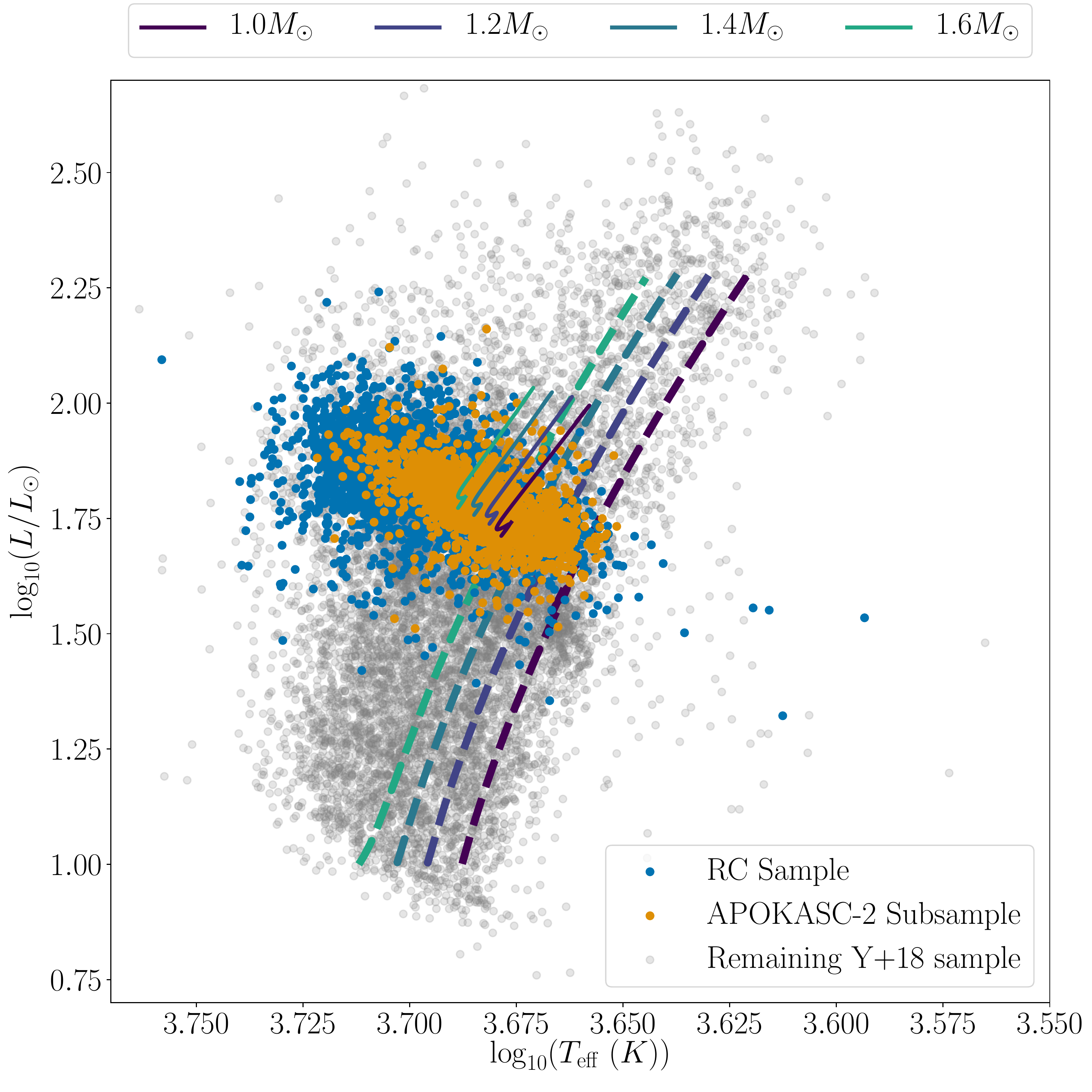}
\caption{HR diagram illustrating the data in our final set of \nstars stars overlaid on the \protect\citetalias{art:yu+2018} sample, along with evolutionary tracks from MESA \protect\citep{art:paxton+2011,art:paxton+2013,art:paxton+2015}\citep[for details about the physical inputs of the models see][]{art:khan+2018}. The stars in the \protect\citetalias{art:yu+2018} sample not in our final selection are in grey. Plotted on top in blue are the stars that in our final sample where the subsample of stars with temperatures reported in APOKASC-2 \protect\citep{art:pinsonneault+2018} are shown in orange. Evolutionary tracks are plotted for for masses ranging between 1.0 and 1.6 solar masses for a metallicity of $Z = 0.01108$ and helium content of $Y = 0.25971$. The dashed lines indicate the Red Giant Branch, whereas the solid lines indicate the main Core Helium Burning stage  of the tracks (the Helium flash (and subflashes) are not included).}
\label{fig:datafig}
\end{figure}

It should be noted that our classification does not specifically account for low-mass, low-metallicity horizontal branch stars, which are therefore expected to be retained in our sample, but are not expected to significantly affect the result as they have similar luminosities to the RC, and no extensive horizontal structure is present on the HR diagram of the \citetalias{art:yu+2018} catalogue, or our subsample thereof (see Figure \ref{fig:datafig}). A fraction of the newly classified stars had masses reported in \citetalias{art:yu+2018} as much higher than we would expect for a RC star. In order to exclude these from our sample, we apply a liberal cut for clump-corrected seismic masses of over $2.2\ M_\odot$, excluding 92 stars from our sample.

To obtain our astrometric sample, we cross-matched the \new{RC stars we selected from the \citetalias{art:yu+2018} sample} with the \gaia DR2 sample\footnote{We make use of the of the  \hyperlink{gaia-kelper.fun}{https://www.gaia-kepler.fun} crossmatch database created by Megan Bedell for this purpose.} \citep{art:gaiacollaboration+2016, art:gaiacollaboration+2018}. In cases of duplicate sources for a given KIC, we selected the star with the lowest angular separation to the target. We did not apply any truncation of the sample based on parallax uncertainty or negative parallax, since this is known to introduce a parallax dependent bias \citep{art:luri+2018}. 

The parallaxes ($\hat{\varpi}$) and parallax uncertainties ($\sigma_{\hat{\varpi}}$) make up our astrometric set of observables. We obtained the apparent magnitudes ($\hat{m}$) and their uncertainties ($\sigma_{\hat{m}}$) from the 2MASS survey for the \Ks band \citep{art:skrutskie+2006} and \gaia DR2 for the \gaia $G$ band, and removed stars that do not have photometry or uncertainties on magnitude in 2MASS. \new{Comparing the magnitude zero-points for the \gaia $G$, $G_{\rm{BP}}$ and $G_{\rm{RP}}$ bands, \cite{art:casagrande+vandenberg2018} found indication of a magnitude-dependent zero-point offset in the \gaia $G$ band magnitudes in the range of $6\ \rm{mag} \lesssim G \lesssim 16.5\ \rm{mag}$, corrected as}

\begin{equation}
G^{\rm corr} = 0.0505 + 0.9966\ G\ ,
\end{equation}

\noindent where $G$ is our uncorrected \gaia $G$ band magnitude. This correction is small, and corresponds to $30\ \rm mmag$ over $10$ magnitudes. We gave all our $G$ band magnitudes a generous uncertainty of $10\ \rm mmag$, the typical uncertainty quoted in \cite{art:gaiacollaboration+2018} for $G = 20$, in order to account for any additional uncertainty incurred by the above correction. \up{It should be noted that a similar relation for the correction of $G$ band magnitudes is presented in \cite{art:maizapellaniz+weiler2018}. This correction places magnitudes about $30\ \rm mmag$ higher than when using the \cite{art:casagrande+vandenberg2018} correction in the applicable magnitude range. We expect the scale of this systematic offset to have a negligible impact on our results, and therefore adopt the \cite{art:casagrande+vandenberg2018} correction in this work for consistency with our chosen $G$ band extinction coefficients and bolometric corrections.}

Our model also uses an extinction for each star in each band. Reddening values are taken from the \cite{art:green+2018} three-dimensional dustmap under the assumption that the distance to the object is that given by \cite{art:bailer-jones+2018}. We note that this is not expected to bias our results towards a previous measure of distance, because the spread in the obtained reddening values, regardless of choice of distance value, falls well within the spread of the prior set on these values in our model. We converted reddening to the band-specific extinction $\hat{A}_\lambda$ using extinction coefficients unique to the \cite{art:green+2018} map for the \Ks band\footnote{These coefficients can be found with the \cite{art:green+2018} \href{http://argonaut.skymaps.info/usage\#units}{usage notes}.}. For the \gaia $G$-band we calculated our band-specific extinction using the mean extinction coefficient presented in \cite{art:casagrande+vandenberg2018}, after converting our reddening value to a measure of $E(B-V)$ following the conventions presented in \cite{art:green+2018}.

The final sample contains \nstars RC stars, with minimal contamination from the 2CL or the RGB, and covers a magnitude range of $\sim8$ to $\sim16\ \rm{mag}$ in $G$ and $\sim6$ to $\sim14\ \rm{mag}$ in $K$. \up{Note that for this magnitude range we expect the \gaia DR2 catalogue to be practically complete, and do not need to apply any selection functions in magnitude.} The data are shown in Figure \ref{fig:datafig} in an HR diagram overlaid on the full \citetalias{art:yu+2018} sample.

\subsection{The APOKASC-2 subsample}
\begin{figure*}
\centering
\includegraphics[width=0.8\textwidth]{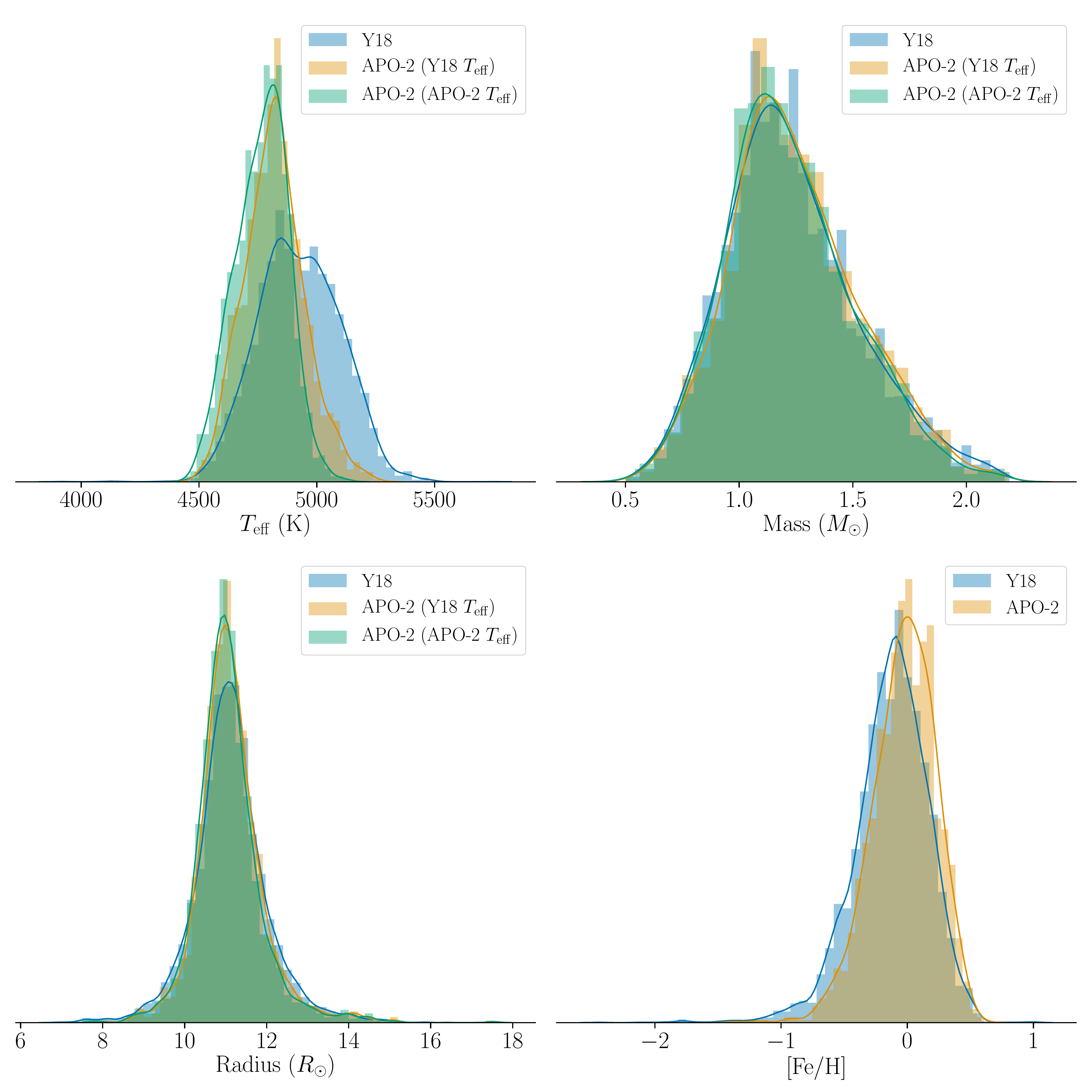}
\caption{Distributions in \teff, mass, radius and \feh of the RC sample \protect\citep{art:yu+2018} and the APOKASC-2 subsample \protect\citep{art:pinsonneault+2014,art:pinsonneault+2018}. In green are the distribution of the APOKASC-2 temperatures, which are overall lower, and the distributions in mass and radius calculated through the direct method for these temperatures. In the labels, `APO-2' represents a shorthand for APOKASC-2.}
\label{fig:apo2}
\end{figure*}

\new{We used temperatures from \cite{art:mathur+2017}, a catalogue compiling temperatures from a diverse set of papers including work with spectroscopy, photometry, and some asteroseismology. In order to investigate the impact of using differing temperature sources on our results, we also included runs on a subsample of 1637 stars that had \teff values reported in the APOKASC-2 catalogue \citep{art:pinsonneault+2014,art:pinsonneault+2018}. When calculating seismic properties from these data, we only changed the values for \teff to our new APOKASC-2 values. In Figure \ref{fig:apo2} we compare the distributions in \teff, mass, radius and \feh of the \citetalias{art:yu+2018} RC sample and the APOKASC-2 subsample. Also shown is the distribution of the APOKASC-2 temperatures, which are overall lower than the \citetalias{art:yu+2018} temperatures, and the distributions in mass and radius calculated through the direct method for these temperatures. Overall the APOKASC-2 subsample represents a lower temperature population, with its most distinct difference being in \teff and \feh.}

\subsection{Obtaining the seismic sample}
The two global observable seismic parameters, \numax and \dnu, scale with fundamental stellar properties as \citep{art:brown+1991, art:kjeldsen+bedding1995}:

\begin{equation}
\frac{\nu_{\rm max}}{\nu_{\rm{max\odot}}} \simeq \left(\frac{M}{M_\odot}\right)\left(\frac{R}{R_\odot}\right)^{-2}\left(\frac{T_{\rm{eff}}}{T_{\rm{eff\odot}}}\right)^{-1/2}\ \rm{and}
\end{equation}

\begin{equation}
\frac{\Delta\nu}{\Delta\nu_\odot} \simeq \left(\frac{M}{M_\odot}\right)^{1/2}\left(\frac{R}{R_\odot}\right)^{-3/2}\ ,
\end{equation}

\noindent where $M$ is the stellar mass, $R$ is the radius, $T_{\rm{eff}}$ is the effective temperature, and $\odot$ indicates a solar value. In this work we used $\nu_{\rm{max\odot}} = 3090\ \pm \ 30\  \mu\rm{Hz}$, $\Delta\nu_\odot = 135.1\ \pm \ 0.1\  \mu\rm{Hz}$ and $T_{\rm{eff\odot}} = 5777$ K \citep{art:huber+2011a}. By rearranging these scaling relations, we can obtain stellar surface gravity and radius as

\begin{equation}\label{eq:logg}
\frac{g}{g_\odot} \simeq \frac{\nu_{\rm{max}}}{\nu_{\rm{max}\odot}} \left(\frac{T_{\rm{eff}}}{T_{\rm{eff}\odot}}\right)^{1/2}\ \rm{and}
\end{equation}

\begin{equation}\label{eq:radius}
\frac{R}{R_\odot} \simeq \left(\frac{\nu_{\rm max}}{\nu_{\rm{max\odot}}}\right) \left(\frac{\Delta\nu}{f_{\Delta\nu}\Delta\nu_\odot}\right)^{-2} \left(\frac{T_{\rm{eff}}}{T_{\rm{eff\odot}}}\right)^{1/2}\ ,
\end{equation}

\noindent where the new term $f_{\Delta\nu}$ is a correction to the \dnu scaling relation in the notation of \cite{art:sharma+2016}. We calculated \fdnu as a function of \feh, \teff, \numax, \dnu and evolutionary state using interpolation in a grid of models \citep{art:sharma+stello2016}. \up{For each perturbation of \teff we recalculated \fdnu, changing no other parameters. We only extracted the correction values \fdnu from the models, and used the seismic parameters and temperature values from our original set, and not the results for these values returned from the grids, in the rest of this work.} We did not include corrections for the \numax scaling relation, because these are more difficult to obtain theoretically \citep{art:belkacem+2011}, and are probably negligible \citep{art:brogaard+2018}. Note that \cite{art:brogaard+2018} found that using corrections by \cite{art:rodrigues+2017} delivers on average slightly smaller stellar properties than using \cite{art:sharma+stello2016} \up{due to differences in how these methods treat the solar surface effect}. Since we used a wide range of bolometric corrections for various temperature perturbations, the method by \cite{art:rodrigues+2017} would be too computationally expensive, and we thus elected to use \cite{art:sharma+stello2016}, \up{which may lead to differences of the order of $\sim 2\%$ in radius than if we had used \cite{art:rodrigues+2017} \citep{art:white+2011}. We discuss the impact of this on our work in Section \ref{sec:discussion}.}

In order to obtain absolute magnitudes for our sample, we used \teff and seismic radii, calculated using \dnu and \numax from the \citetalias{art:yu+2018} catalogue through equation (\ref{eq:radius}), to calculate the stellar luminosity as

\begin{equation}
L_* = 4\pi\sigma_{\rm sb}R^2T_{\rm eff}^4\ .
\end{equation}

\noindent Here $L_*$ is the luminosity of the star and $\sigma_{\rm{sb}}$ is the Stefan-Boltzmann constant. This was converted to a bolometric magnitude as in \cite{art:casagrande+vandenberg2014}:

\begin{equation}
M_{\rm bol} = -2.5\log_{10}(L_* / L_\odot) + M_{\rm bol \odot}\ ,
\end{equation}


\noindent where $L_\odot$ is the solar luminosity, and we have adopted $M_{\rm bol \odot} = 4.75$ \citep{art:casagrande+vandenberg2014,art:casagrande+vandenberg2018b,art:casagrande+vandenberg2018}. We calculated the bolometric correction (BC) in the 2MASS $K$ and \gaia $G$ bands with the method described by \cite{art:casagrande+vandenberg2014,art:casagrande+vandenberg2018b,art:casagrande+vandenberg2018} using \teff, \feh and \logg, and without accounting for extinction. \up{Since we are using a distance-independent measure of luminosity to calculate an absolute magnitude}, accounting for this in the BC would bias our results. Because our method requires tweaking our values for \teff, we recalculated the \logg used to find the BC through the scaling relation in equation (\ref{eq:logg}), \up{as well as our values for \fdnu,} for each different set of temperatures, and thus obtained a full set of bolometric corrections \up{and corrections to the scaling relations} for each temperature perturbation. Our values of absolute magnitude were then given by

\begin{equation}
\hat{M}_\lambda = M_{\rm bol} - BC_\lambda\ ,
\end{equation}

\noindent where $\lambda$ is the relevant band, $M_{\rm bol}$ is the bolometric luminosity and $BC_\lambda$ is the bolometric correction in that band. Uncertainties on $\hat{M}_\lambda$ were propagated through from the uncertainties on seismic parameters and effective temperatures, including those on the solar seismic parameters. Uncertainties on the BCs were estimated using a Monte Carlo method with 5000 iterations for 1000 randomly selected stars from our sample. We found an uncertainty of $0.3\ \rm mag$ for all BCs in the $G$ band. For the $K$ band we found $0.05\ \rm mag$ for stars with a fractional temperature uncertainty of $< 2.5\%$, and $0.09\ \rm mag$ for those with larger fractional uncertainties on temperature. We discuss the systematic uncertainties on \fdnu in Section \ref{sec:discussion}.

\section{Locating the Red Clump using hierarchical Bayesian modelling} \label{sec:method}
In order to test systematics \up{in asteroseismology and \gaia} using the Red Clump (RC), \up{we aim to find the location and spread of the RC in absolute magnitude using both sets of data separately.} To obtain these RC parameters, we fitted a model for the distribution of RC stars in `true' absolute magnitude, either inferred from an observed absolute magnitude (asteroseismic) or inferred from apparent magnitude, parallax, and extinction (astrometric).

We \nnew{built a pair of} Bayesian hierarchical models with latent parameters that allow us to infer key values such as the distance and the true absolute magnitude from the data and the model. The latent parameters form a stepping stone between our population model, which is described by \textit{hyperparameters}, and the observations. We use a latent parameter for each star to infer the `true' distribution of the absolute magnitudes, while fitting our population level model to these inferred `true' absolute magnitudes, instead of to the observations themselves. Many aspects of our hierarchical models, especially those for the \gaia data, are similar to those used for the same purpose by \citetalias{art:hawkins+2017} with some improvements.

To fit to the position and spread of RC stars while also isolating any outlier contaminants, we applied the mixture model \citep{art:hogg+2010} utilised by \citetalias{art:hawkins+2017}. In this case, we employed two generative models weighted by the mixture-model weighting factor $Q$. For these we used two normal distributions: one for the inlier population of RC stars, with a mean $\mu_{\rm RC}$ and a \new{standard deviation (spread)} $\sigma_{\rm RC}$, and a broad outlier distribution centered in the same location ($\mu_{\rm RC})$ but with a spread of $\sigma_o$, which must always be larger than $\sigma_{\rm RC}$. The likelihood to obtain an absolute magnitude $M_i$ given this mixture model is then

\begin{equation}\label{eq:pMi}
\begin{split}
p(&M_i | \theta_{\rm RC}) \\
&=  Q\mathcal{N}(M_i | \mu_{\rm RC}, \sigma_{\rm RC}) +  (1-Q)\mathcal{N}(M_i | \mu_{\rm RC}, \sigma_{\rm o})\ ,
\end{split}
\end{equation}

\noindent where $M_i$ is the true absolute magnitude for a given datum $i$, $\theta_{RC} = \{\mu_{\rm RC}, \sigma_{\rm RC}, Q, \sigma_{\rm o}\}$\ are the model hyperparameters (which inform the population of latent parameters) and $\mathcal{N}(x|\mu, \sigma)$ represents a normal distribution evaluated at $x$, with a mean $\mu$ and a spread $\sigma$.\footnote{Note that the spread $\sigma$ as listed in $\mathcal{N}(x|\mu, \sigma)$ is not a variance, but a standard deviation, since we are following the nomenclature used in \texttt{pystan}.}

\subsection{The asteroseismic model}
For our asteroseismic model, we used a calculated measure of the absolute magnitude ($\hat{M}$) from asteroseismology, along with appropriate uncertainties ($\sigma_{\hat{M}}$), as our data. We used a latent variable model to infer the true value of the absolute magnitude. Given our data and the hyperparameters on our mixture model $\theta_{\rm RC}$, we can use Bayes' theorem to find the unnormalised posterior probability of our model:

\begin{equation}\label{eq:postprob_ast}
\begin{split}
p(\theta_{\rm RC} | \mathcal{D}) \propto  p(\theta_{\rm RC}) \prod_{i=1}^{\rm N} p(\mathcal{D}_i | M_i)p(M_i | \theta_{\rm RC})\ .
\end{split}
\end{equation}

\noindent Here, N is the number of points in our data set $\mathcal{D} = \{\hat{M}, \sigma_{\hat{M}}\}$, $p(\mathcal{D}_i | M_i)$ is our likelihood function, $p(\theta_{\rm RC})$ represents the priors on the hyperparameters, and $p(M_i | \theta_{\rm RC})$ is the probability to obtain our latent parameters \up{(the true absolute magnitudes)} given our hyperparameters. 

\begin{figure}
\centering
\includegraphics[width=.48\textwidth]{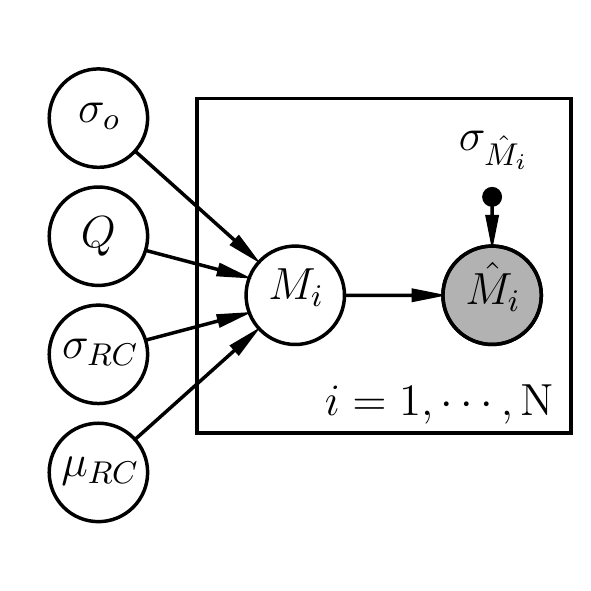}
\caption{An probabilistic graphical model of the asteroseismic model, represented algebraically in equation \ref{eq:postprob_ast}. Shaded circles indicate observed data, \new{whereas solid black circles represent fixed parameters}, such as the uncertainty on the observed data. The hyperparameters $\theta_{\rm RC}$ can be seen on the left, and inform the set of latent parameters $M_i$, which in turn relate to the observed data $\hat{M_i}$ and $\sigma_{\hat{M}_i}$. N is the number of data points in our sample.}
\label{fig:asteromodel}
\end{figure}

The likelihood to obtain our data given our parameters is then

\begin{equation}
p(\mathcal{D}_i | M_i) = \mathcal{N}(\hat{M}_i | M_i, \sigma_{\hat{M}_i})\ ,
\end{equation}

\noindent where $M_i$ is the true absolute magnitude. Here, $M_i$ is a latent parameter that is drawn from from the likelihood function $p(M_i | \theta_{\rm RC})$ (equation \ref{eq:pMi}), to which our hyperparameters are fitted. A probabilistic graphical model of the asteroseismic model is shown in Figure \ref{fig:asteromodel}.

\subsection{The astrometric model}
Fitting the absolute magnitude for the \gaia DR2 sample required a more involved approach, since we wanted to work directly with parallax \citep{art:luri+2018}. We used a set of three latent parameters, $\alpha_i = \{M_i, r_i, A_i\}$, where $M_i$ is the absolute magnitude in a given band, $r_i$ is the distance and $A_i$ is the extinction in a given band. We also include two additional hyperparameters: \oozp, the parallax zero-point offset and $L$, the length scale of the exponentially decreasing space density prior on distance \citep{art:astraatmadja+bailer-jones2016,art:astraatmadja+bailer-jones2016a, art:astraatmadja+bailer-jones2017}. This prior, which is necessary to treat negative parallax values, has already successfully been applied to \gaia DR2 data \citep{art:bailer-jones+2018} and its use is recommended for this purpose within the \gaia DR2 release papers \citep{art:luri+2018}.

Some extra care was also required in the treatment of parallax uncertainties for this sample. \cite{art:lindegren+2018} found parallaxes to be correlated on scales below $40^\circ$, with increasing strength at smaller separations, and quantified their covariance using quasar parallaxes. They found the positive covariance $V_\varpi$ for these scales to be reasonably approximated by the fitted relation

\begin{equation}\label{eq:covariance}
V_\varpi(\theta) \simeq (285\ \mu \rm as^2) \times \exp(-\theta/14^\circ)\ ,
\end{equation}

\noindent where $\theta$ is the angular separation between two targets in degrees. The fit corresponds to a RMS amplitude of $\sqrt{285\ \mu \rm as^2} \approx 17\ \mu\rm as$. This relation was recently applied by \cite{art:zinn+2018}, who found that the \cite{art:lindegren+2018} relation resulted in the best goodness-of-fit for their models of the parallax zero-point offset, over both a similar relation by \cite{art:zinn+2017} based on TGAS data, and not including parallax covariances altogether.

We generated a covariance matrix $\underline{\Sigma}$ for our sample:

\begin{equation}\label{eq:covar}
\Sigma_{ij} = V_\varpi(\theta_{ij}) + \delta_{ij}\sigma_{\hat{\varpi}_i}\sigma_{\hat{\varpi}_j}\ ,
\end{equation}

\noindent where $\theta_{ij}$ is the angular separation between stars $i$ and $j$, and $\delta_{ij}$ is the Kronecker delta function. 

Given these new additions, our set of data was $\mathcal{D} = \{\hat{\varpi}, \underline{\Sigma}, \hat{m}, \sigma_{\hat{m}}, \hat{A} \}$, where all symbols are as defined above and $\hat{A}$ is the band specific extinction. We can use Bayes' theorem, as before, to find the unnormalised posterior probability of our model as

\begin{equation}
\begin{split}
p(&\theta_{\rm RC}, \varpi_{\rm zp}, L, \alpha | \mathcal{D}) \\
&\propto  p(\theta_{\rm RC}, \varpi_{\rm zp}, L, \alpha)\ p(\mathcal{D} | \theta_{\rm{RC}}, \varpi_{\rm{zp}}, L, \alpha)\ ,
\end{split}
\end{equation}

\noindent where $p(\mathcal{D} | \theta_{\rm{RC}}, \varpi_{\rm{zp}}, L, \alpha)$ is now our likelihood function and $p(\theta_{\rm RC}, \varpi_{\rm zp}, L,  \alpha)$ represents the priors on our hyper- and latent parameters. Our likelihood function relates to two observables as, 

\begin{equation}\label{eq:like_gaia}
p(\mathcal{D} | \theta_{\rm{RC}}, \varpi_{\rm{zp}}, L, \alpha) = p(\hat{\varpi} | r, \oozp, \underline{\Sigma}) \times p(\hat{m} | \alpha, \sigma_{\hat{m}})\ .
\end{equation}

\noindent Note that the parallax only depends on the latent parameter for distance, $r$. Since parallax values are correlated, $p(\hat{\varpi} | r, \oozp, \underline{\Sigma})$ was evaluated for all data simultaneously, whereas $p(\hat{m} | \alpha, \sigma_{\hat{m}})$ was evaluated at every datum $i$. This means that our full posterior probability takes the form

\begin{equation}\label{eq:postprob_gaia}
\begin{split}
p(&\theta_{\rm RC}, \varpi_{\rm zp}, L, \alpha | \mathcal{D}) \\
&\propto  p(\theta_{\rm RC}, \varpi_{\rm zp}, L)\ p(\hat{\varpi} | r, \oozp, \underline{\Sigma})\ \times \\
&\prod_{i=1}^{\rm N} p(\hat{m}_i | \alpha_i, \sigma_{\hat{m}_i})\ p(\alpha_i | \theta_{\rm RC}, \varpi_{\rm zp}, L)\ ,
\end{split}
\end{equation}

\noindent where the first term represents the priors on our hyperparemeters, the second term is the likelihood to obtain our observed parallaxes, the third is the likelihood to obtain an observed magnitude, and the fourth gives the probability to obtain the latent parameters, given the hyperparameters.\\

The second component of equation \ref{eq:postprob_gaia} is the probability of obtaining the observed parallax given our latent parameters and our covariance matrix. Since we treated our parallax uncertainties as correlated, we evaluated these probabilities for the full set using a multivariate normal distribution:

\begin{equation}\label{parallax_like}
p(\hat{\varpi} | r, \oozp, \underline{\Sigma}) = \mathcal{N}(\hat{\varpi} | 1/r + \oozp , \underline{\Sigma})\ ,
\end{equation}

\noindent where $1/r$ defines the \emph{true} parallax. The latent parameters for the distance $r_i$ were drawn from an exponentially decreasing space density prior \citep{art:bailer-jones2015}, which goes as

\begin{equation}\label{eq:bjprior}
p(r_i | L) = \frac{1}{2L^3}r_i^2\exp(-r_i/L)\ ,
\end{equation}

\noindent and thus depends on the length scale hyperparameter $L$. This prior has a mode at $2L$, beyond which it decreases exponentially.

The third component of equation \ref{eq:postprob_gaia} is then 

\begin{equation}\label{eq:magnitude_like}
p(\hat{m}_i | \alpha_i, \sigma_{\hat{m}_i}) = \mathcal{N}(\hat{m}_i | m_i, \sigma_{\hat{m_i}})\ ,
\end{equation}

\noindent where $m_i$ is the \textit{true} apparent magnitude, and is drawn from the relation

\begin{equation}\label{eq:magnitudeequation}
m_i = M_i + 5\logten(r_i) - 5 + A_i\ .
\end{equation}

\begin{figure}
\centering
\includegraphics[width=.48\textwidth]{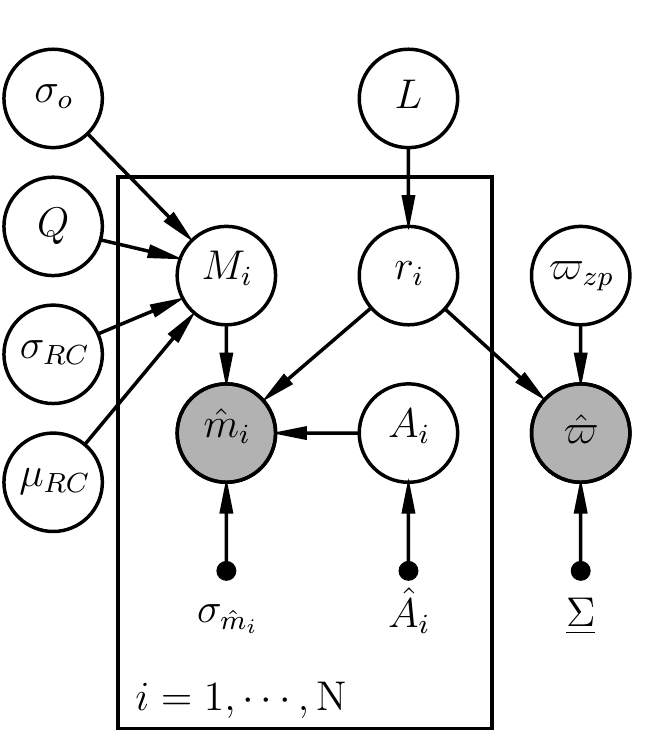}
\caption{An acyclic diagram of the astrometric model, represented algebraically in equation \ref{eq:postprob_gaia}. Conventions are the same as for Figure \ref{fig:asteromodel}. The full parallax covariance matrix is denoted as $\underline{\Sigma}$; it should be noted that the parallax likelihood is evaluated across the full set as a multivariate normal distribution.}
\label{fig:gaiamodel}
\end{figure}

\noindent Here, we have used the inferred true values for absolute magnitude, distance and extinction to calculate apparent magnitude. As for the seismic method, the true absolute magnitude $M_i$ was drawn from the likelihood $p(M_i | \theta_{RC})$, as given in equation \ref{eq:pMi}. The final latent parameter $A_i$ is given a prior as

\begin{equation}
p(A_i | \hat{A}_i) = \mathcal{N}(A_i | \hat{A}_i, 0.05)\ ,
\end{equation}

\noindent a normal distribution with a spread of $0.05\ \rm{mag}$, where $\hat{A}_i$ is our observed value for the extinction \citep{art:green+2018}. A probabilistic graphical model of the astrometric model is shown in Figure \ref{fig:gaiamodel}.

\subsection{Priors on the hyperparameters}
The priors on the hyperparameters were, where possible, identical across both models. For the asteroseismic model, our priors took the form of

\begin{equation}\label{eq:asteropriors}
\begin{split}
\mu_{\rm RC} &\sim \mathcal{N}(\mu_{\rm H}, 1)\\
\sigma_{\rm RC} &\sim \mathcal{N}(0, 1)\\
Q &\sim\mathcal{N}(1, 0.25)\\
\sigma_{\rm o} &\sim \mathcal{N}(3, 2)\ , \\
\end{split}
\end{equation}

\noindent where $\mu_{\rm H}$ is the absolute magnitude of the RC in the relevant passband, as reported by \citetalias{art:hawkins+2017}, and $\sigma_{\rm RC}$ must be above 0. It should be noted that, in order to evaluate the hierarchical mixture model in \texttt{PyStan}, $\sigma_{\rm o}$ is expressed in units of $\sigma_{\rm RC}$ and must always be larger than 1 to ensure the two components of the mixture model do not switch roles. $Q$ must fall within the range 0.5 to 1, because we expect an inlier-dominated sample.

For the astrometric method, we introduced the two new parameters $\varpi_{\rm zp}$ and $L$, and applied a new prior to $\mu_{\rm RC}$ and \sigrc, while the priors for the other hyperparameters remained the same:

\begin{equation}\label{eq:gaiapriors}
\begin{split}
\mu_{\rm RC} &\sim \mathcal{N}(\mu_{\rm RC, seis}, \sigma_{\mu_{\rm RC, seis}})\\
\sigma_{\rm RC} &\sim \mathcal{N}(\sigma_{\rm RC, seis}, \sigma_{\sigma_{\rm RC, seis}})\\
L &\sim \mathcal{U}(0.1, 4000)\\
\oozp &\sim \mathcal{N}(0, 500)\ .\\
\end{split}
\end{equation}

\noindent Here, $\mathcal{U}$ denotes a uniform distribution with the lower and upper limits as arguments, and the units of $\varpi_{\rm zp}$ and $L$ are $\mu \rm as$ and $\rm kpc$, respectively. The quantities $\mu_{\rm RC, seis}$ and $\sigma_{\rm RC, seis}$ are the medians of the posterior distributions on $\mu_{\rm RC}$ and \sigrc from the asteroseismic model, and $\sigma_{\mu_{\rm RC, seis}}$ and $\sigma_{\sigma_{\rm RC, seis}}$ are the spreads on the posteriors, effectively allowing us to explore what value of the parallax-zero point offset, $\oozp$, recovers the results we see using asteroseismology.

\new{Finally, for runs where we investigated the impact of literature values for \oozp on our RC parameters, we set the priors on \murc and \sigrc to those used on our seismic run, and applied a prior on \oozp as}

\begin{equation} \label{eq:zeropointpriors}
\varpi_{\rm zp} \sim \mathcal{N}(\varpi_{\rm zp, lit}, \sigma_{\varpi_{\rm zp, lit}})\ .
\end{equation}

\new{\noindent Here, $\varpi_{\rm zp, lit}$ and $\sigma_{\varpi_{\rm zp, lit}}$ are values and uncertainties on said values from the literature.}\\

We drew samples from the posterior distributions using \texttt{PyStan} version \texttt{2.18.0.0}, with four chains and 5000 iterations, with half of the iterations used as burn-in. \nnew{Appropriate convergence of our chains was evaluated using the $\hat{R}$ diagnostic.} \footnote{Our code is open and can be found on Github at \url{https://www.github.com/ojhall94/halletal2019}} 

\section{Results} \label{sec:results}
\subsection{Results from asteroseismology}
To see how the absolute magnitude $\mu_{\rm RC}$ and spread $\sigma_{\rm RC}$ of the RC change given our input data, we applied two changes to calculations for seismic absolute magnitude. First, we perturbed the temperature by a value $\Delta T_{\rm eff}$ that ranged between $-50$ and $50\ \rm{K}$, in steps of $10\ \rm{K}$. Second, we propagated these temperatures, along with the original and unperturbed uncertainties on \teff, \numax and \dnu, through the seismic scaling relations to find luminosity. We did this both with and without calibrations for the \dnu scaling relation obtained by the grid interpolation method by \cite{art:sharma+2016}. The perturbed temperatures were also used in the grid interpolation required to obtain the correction \citep{art:sharma+stello2016}, and the corrections were thus recalculated for each change in temperature. We also calculated BCs for each set of temperatures, and recalculated a seismic \logg given the perturbed temperatures for each calculation of the BCs \citep{art:casagrande+vandenberg2014, art:casagrande+vandenberg2018b, art:casagrande+vandenberg2018}. \up{Seismic radii were calculated per equation \ref{eq:radius}, which were in turn used to calculate luminosities and were combined with the BCs to compute our absolute magnitudes, resulting in 22 individual sets that differ in corrections to the seismic scaling relations and temperature scale, for both photometric bands.}

\begin{figure*}
    \centering
    \includegraphics[width=\textwidth]{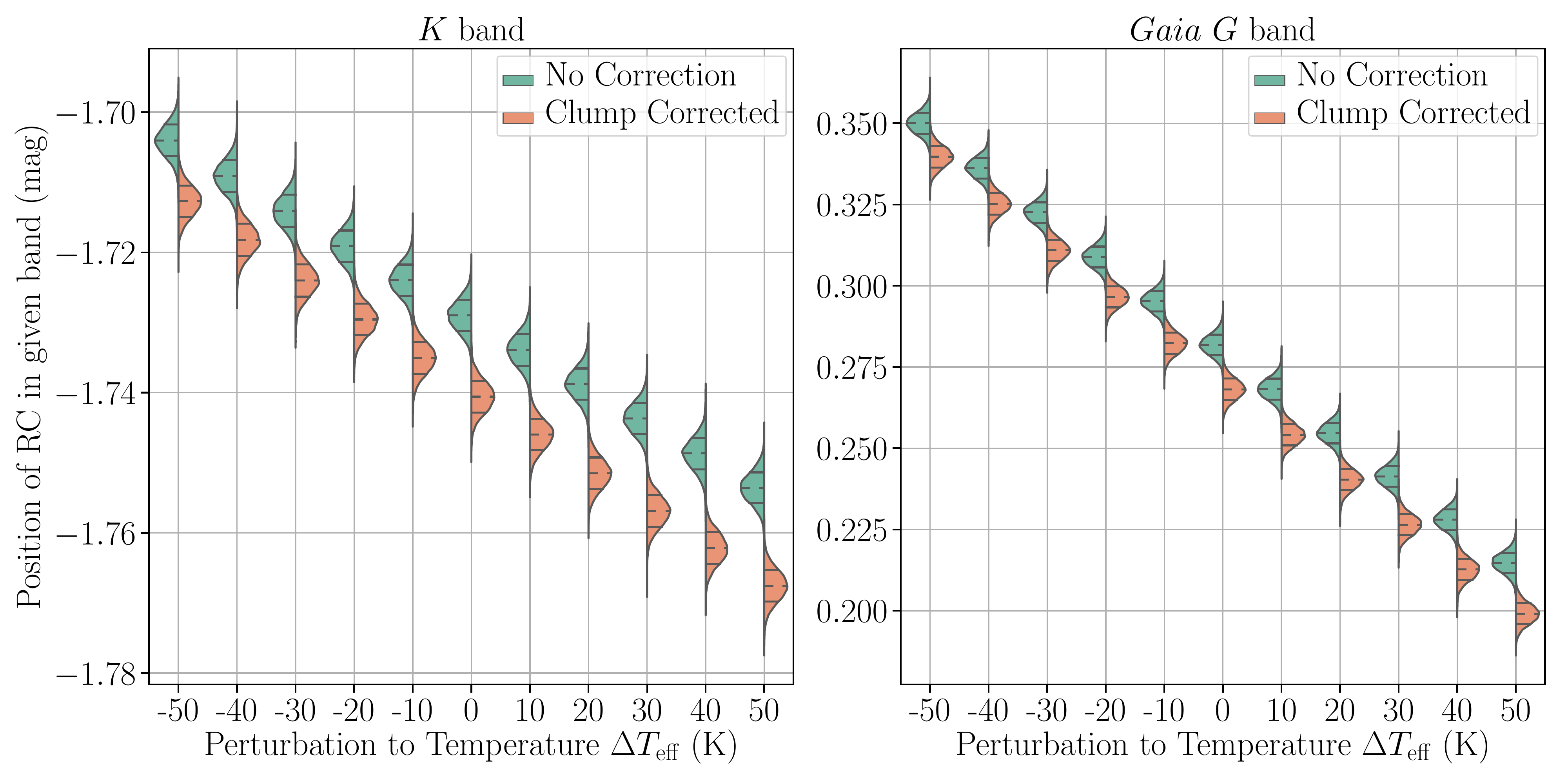}
    \caption{The posterior distributions on the position of the Red Clump in the 2MASS \Ks band (left) and \gaia $G$ band (right), as a function of overall perturbation to the temperature values $\Delta T_{\rm eff}$ using asteroseismology, both with (orange) and without (green) corrections to the \dnu scaling relation \citep{art:sharma+2016}. The dashed horizontal lines indicate the median on the posteriors, and the solid horizontal lines represent \nnew{the 1$\sigma$ credible intervals}. The posteriors' magnitudes along the x-axis are indicative of power with arbitrary units, whereas their shape along the y-axis indicates the spread in the posterior result.}
    \label{fig:yu_posteriors}
\end{figure*}

\up{Our results for our \citetalias{art:yu+2018} sample are shown in Tables \ref{tab:yu_k} \& \ref{tab:yu_g} where} we present the medians of the posterior distributions for our hyperparameters for the 2MASS $K$ band and \gaia $G$ band respectively, both with and without a correction to the \dnu scaling relation, for various changes in temperature scale. Uncertainties are given as the 1$\sigma$ credible intervals. Where the posterior distributions are approximately Gaussian we quote a symmetric single uncertainty. The change of the posterior on the magnitude of the RC $\mu_{\rm RC}$ alone, given the input, can be seen in Figure \ref{fig:yu_posteriors}.

\begin{figure*}
    \centering
    \includegraphics[width=\textwidth]{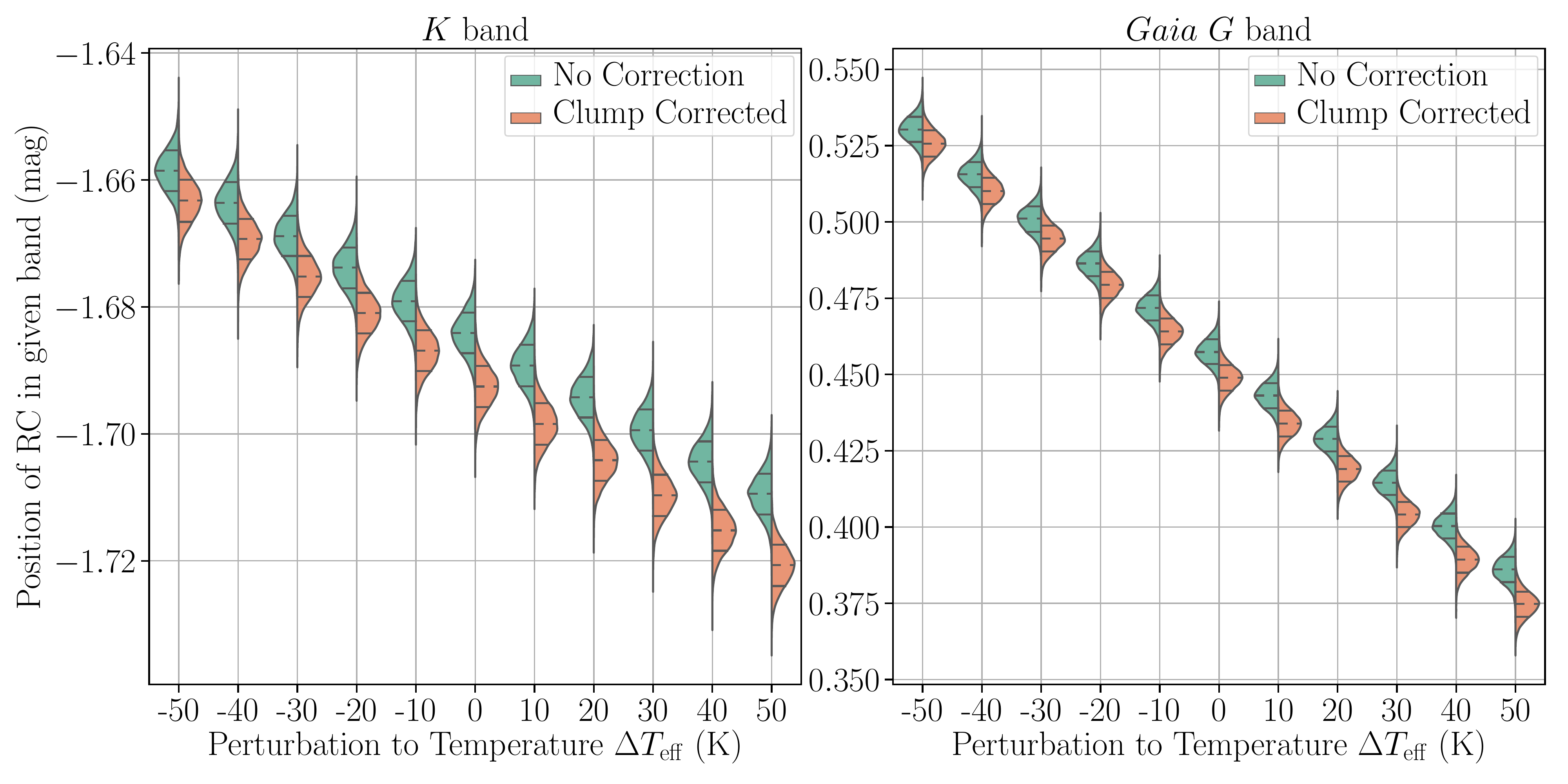}
    \caption{The same as Figure \ref{fig:yu_posteriors}, except using only stars both in our sample and the APOKASC-2 \citep{art:pinsonneault+2018} sample, replacing \teff with those reported in APOKASC-2.}
    \label{fig:apokasc2_posteriors}    
\end{figure*}

\new{For our APOKASC-2 temperature subsample of 1637 stars, we reran our models using the same methodology as before,} simply substituting the temperatures and temperature uncertainties reported in \cite{art:pinsonneault+2018} for those in \citetalias{art:yu+2018} for those stars, and making no other changes. Note that the change in temperature values carried through to the calculation of the bolometric corrections and corrections to the scaling relations for each run. The results of this are presented in Tables \ref{tab:apo_k} and \ref{tab:apo_g} for all hyperparameters, as with the run on the full sample. The change in the posteriors on the position of the RC is shown for this reduced sample in Figure \ref{fig:apokasc2_posteriors}.

\begin{table*}
    \begin{tabular}{rrrrr|rrrr}
    \toprule
    {} & \multicolumn{4}{|c|}{No Correction} & \multicolumn{4}{|c|}{Clump Corrected} \\
    \cmidrule(lr){2-5}\cmidrule(lr){6-9}
    \multicolumn{1}{c}{$\Delta T_{\rm eff}\ (K)$} & \multicolumn{1}{c}{$\mu_{\rm RC}\ (\rm mag)$} & \multicolumn{1}{c}{$\sigma_{\rm RC}\ (\rm mag)$}  & \multicolumn{1}{c}{$Q$} &  \multicolumn{1}{c}{$\sigma_{\rm o}\ (\sigma_{\rm RC})$} & \multicolumn{1}{c}{$\mu_{\rm RC}\ (\rm mag)$} & \multicolumn{1}{c}{$\sigma_{\rm RC}\ (\rm mag)$}  & \multicolumn{1}{c}{$Q$} &  \multicolumn{1}{c}{$\sigma_{\rm o}\ (\sigma_{\rm RC})$} \\
    \midrule
        \textbf{-50.0} & -1.704 $\pm$ 0.002 &   0.03 $\pm$ 0.003 &  0.92 $\pm$ 0.01 &  10.35$_{-1.01}^{+1.17}$    & -1.713 $\pm$ 0.002 &   0.034 $\pm$ 0.004 &  0.91 $\pm$ 0.01 &   8.85$_{-0.93}^{+1.09}$\\
        \textbf{-40.0} & -1.709 $\pm$ 0.002 &   0.03 $\pm$ 0.003 &  0.92 $\pm$ 0.01 &  10.33$_{-1.01}^{+1.22}$    & -1.718 $\pm$ 0.002 &   0.033 $\pm$ 0.004 &  0.91 $\pm$ 0.01 &   9.11$_{-1.04}^{+1.12}$ \\
        \textbf{-30.0} & -1.714 $\pm$ 0.002 &   0.03 $\pm$ 0.003 &  0.92 $\pm$ 0.01 &   10.4$_{-1.04}^{+1.15}$    & -1.724 $\pm$ 0.002 &   0.033 $\pm$ 0.004 &  0.91 $\pm$ 0.01 &   9.16$_{-0.96}^{+1.12}$\\
        \textbf{-20.0} & -1.719 $\pm$ 0.002 &  0.029 $\pm$ 0.003 &  0.92 $\pm$ 0.01 &  10.55$_{-1.05}^{+1.15}$    &  -1.73 $\pm$ 0.002 &   0.033 $\pm$ 0.004 &  0.91 $\pm$ 0.01 &   9.22$_{-0.91}^{+1.05}$ \\
        \textbf{-10.0} & -1.724 $\pm$ 0.002 &   0.03 $\pm$ 0.003 &  0.92 $\pm$ 0.01 &  10.49$_{-1.03}^{+1.13}$    & -1.735 $\pm$ 0.002 &   0.033 $\pm$ 0.004 &  0.91 $\pm$ 0.01 &   9.16$_{-0.98}^{+1.09}$\\
        \textbf{0.0  } & -1.729 $\pm$ 0.002 &   0.03 $\pm$ 0.003 &  0.92 $\pm$ 0.01 &  10.33$_{-1.01}^{+1.19}$    & -1.741 $\pm$ 0.002 &   0.033 $\pm$ 0.004 &  0.91 $\pm$ 0.01 &   9.18$_{-0.92}^{+1.09}$ \\
        \textbf{10.0 } & -1.734 $\pm$ 0.002 &  0.029 $\pm$ 0.003 &  0.92 $\pm$ 0.01 &  10.44$_{-0.97}^{+1.07}$    & -1.746 $\pm$ 0.002 &   0.032 $\pm$ 0.004 &  0.91 $\pm$ 0.01 &   9.36$_{-1.05}^{+ 1.2}$\\
        \textbf{20.0 } & -1.739 $\pm$ 0.002 &   0.03 $\pm$ 0.004 &  0.92 $\pm$ 0.01 &  10.32$_{-1.02}^{+1.17}$    & -1.752 $\pm$ 0.002 &   0.033 $\pm$ 0.004 &  0.91 $\pm$ 0.01 &   9.19$_{-1.01}^{+1.16}$ \\
        \textbf{30.0 } & -1.744 $\pm$ 0.002 &   0.03 $\pm$ 0.003 &  0.92 $\pm$ 0.01 &  10.41$_{-0.99}^{+1.06}$    & -1.757 $\pm$ 0.002 &   0.032 $\pm$ 0.004 &  0.91 $\pm$ 0.01 &   9.37$_{-1.02}^{+1.16}$\\
        \textbf{40.0 } & -1.749 $\pm$ 0.002 &   0.03 $\pm$ 0.003 &  0.92 $\pm$ 0.01 &  10.41$_{-1.02}^{+1.18}$    & -1.762 $\pm$ 0.002 &   0.032 $\pm$ 0.004 &  0.91 $\pm$ 0.01 &   9.37$_{-0.97}^{+1.14}$ \\
        \textbf{50.0 } & -1.754 $\pm$ 0.002 &   0.03 $\pm$ 0.003 &  0.92 $\pm$ 0.01 &  10.27$_{-1.01}^{+1.12}$    & -1.768 $\pm$ 0.002 &   0.032 $\pm$ 0.004 &  0.91 $\pm$ 0.01 &   9.25$_{-1   }^{+1.16}$ \\
    \bottomrule
    \end{tabular}
\caption{Medians of the posterior distributions for hyperparameters of our seismic model, for the 2MASS $K$ band, for \nstars stars from the \citetalias{art:yu+2018} sample. Uncertainties are taken as the 1$\sigma$ credible intervals, and are listed as a single value for cases where the posterior was approximately Gaussian. Values are listed for data that have been left uncorrected (No Correction) and data with an appropriate correction to the seismic scaling relations (Clump Corrected). $\Delta\teff$ is the global shift to our values of \teff, \murc is the position of the RC in absolute magnitude, \sigrc is the spread of the RC in absolute magnitude, $Q$ is the mixture model weighting factor (and the effective fraction of stars considered inliers), and $\sigma_o$ is the spread of our outlier population, expressed in terms of \sigrc.}
\label{tab:yu_k}
\end{table*}

\begin{table*}
    \begin{tabular}{rrrrr|rrrr}
    \toprule
    {} & \multicolumn{4}{|c|}{No Correction} & \multicolumn{4}{|c|}{Clump Corrected} \\
    \cmidrule(lr){2-5}\cmidrule(lr){6-9}
    \multicolumn{1}{c}{$\Delta T_{\rm eff}\ (K)$} & \multicolumn{1}{c}{$\mu_{\rm RC}\ (\rm mag)$} & \multicolumn{1}{c}{$\sigma_{\rm RC}\ (\rm mag)$}  & \multicolumn{1}{c}{$Q$} &  \multicolumn{1}{c}{$\sigma_{\rm o}\ (\sigma_{\rm RC})$} & \multicolumn{1}{c}{$\mu_{\rm RC}\ (\rm mag)$} & \multicolumn{1}{c}{$\sigma_{\rm RC}\ (\rm mag)$}  & \multicolumn{1}{c}{$Q$} &  \multicolumn{1}{c}{$\sigma_{\rm o}\ (\sigma_{\rm RC})$} \\
    \midrule
        \textbf{-50.0} & -1.659 $\pm$ 0.003 &  0.029 $\pm$ 0.004 &  0.9 $\pm$ 0.02 &   9.2$_{-1.09}^{+ 1.2}$    &  -1.663 $\pm$ 0.003 &   0.031 $\pm$ 0.005 &  0.89 $\pm$ 0.02 &   8.46$_{-1.06}^{+1.19}$   \\
        \textbf{-40.0} & -1.664 $\pm$ 0.003 &  0.029 $\pm$ 0.004 &  0.9 $\pm$ 0.02 &  9.14$_{-1.08}^{+1.18}$    &  -1.669 $\pm$ 0.003 &   0.032 $\pm$ 0.005 &  0.89 $\pm$ 0.02 &    8.4$_{- 1.1}^{+1.16}$    \\
        \textbf{-30.0} & -1.669 $\pm$ 0.003 &  0.029 $\pm$ 0.004 &  0.9 $\pm$ 0.02 &  9.13$_{-1.11}^{+1.23}$    &  -1.675 $\pm$ 0.003 &   0.031 $\pm$ 0.005 &  0.89 $\pm$ 0.02 &   8.53$_{-1.06}^{+1.16}$   \\
        \textbf{-20.0} & -1.674 $\pm$ 0.003 &  0.029 $\pm$ 0.004 &  0.9 $\pm$ 0.02 &  9.15$_{- 1.1}^{+1.26}$    &  -1.681 $\pm$ 0.003 &   0.031 $\pm$ 0.005 &  0.89 $\pm$ 0.02 &   8.43$_{-1.06}^{+1.24}$    \\
        \textbf{-10.0} & -1.679 $\pm$ 0.003 &   0.03 $\pm$ 0.004 &  0.9 $\pm$ 0.02 &  9.11$_{-1.09}^{+1.18}$    &  -1.687 $\pm$ 0.003 &   0.032 $\pm$ 0.005 &  0.89 $\pm$ 0.02 &   8.37$_{-1.11}^{+1.23}$   \\
        \textbf{0.0  } & -1.684 $\pm$ 0.003 &  0.029 $\pm$ 0.004 &  0.9 $\pm$ 0.02 &  9.13$_{- 1.1}^{+1.25}$    &  -1.693 $\pm$ 0.003 &   0.031 $\pm$ 0.005 &  0.89 $\pm$ 0.02 &    8.5$_{-1.08}^{+1.18}$    \\
        \textbf{10.0 } & -1.689 $\pm$ 0.003 &   0.03 $\pm$ 0.004 &  0.9 $\pm$ 0.02 &  9.08$_{-1.08}^{+ 1.2}$    &  -1.698 $\pm$ 0.003 &   0.032 $\pm$ 0.005 &  0.89 $\pm$ 0.02 &   8.41$_{-1.06}^{+ 1.2}$   \\
        \textbf{20.0 } & -1.694 $\pm$ 0.003 &   0.03 $\pm$ 0.004 &  0.9 $\pm$ 0.02 &  9.04$_{-1.06}^{+1.26}$    &  -1.704 $\pm$ 0.003 &   0.032 $\pm$ 0.005 &  0.89 $\pm$ 0.02 &   8.44$_{-1.08}^{+1.21}$    \\
        \textbf{30.0 } & -1.699 $\pm$ 0.003 &  0.029 $\pm$ 0.004 &  0.9 $\pm$ 0.02 &   9.1$_{-1.07}^{+1.17}$    &   -1.71 $\pm$ 0.003 &   0.033 $\pm$ 0.005 &   0.9 $\pm$ 0.02 &   8.29$_{-1.05}^{+1.21}$   \\
        \textbf{40.0 } & -1.704 $\pm$ 0.003 &  0.029 $\pm$ 0.004 &  0.9 $\pm$ 0.02 &  9.12$_{- 1.1}^{+1.18}$    &  -1.715 $\pm$ 0.003 &   0.032 $\pm$ 0.005 &  0.89 $\pm$ 0.02 &   8.43$_{-1.07}^{+1.23}$    \\
        \textbf{50.0 } & -1.709 $\pm$ 0.003 &  0.029 $\pm$ 0.004 &  0.9 $\pm$ 0.02 &  9.15$_{-1.09}^{+1.24}$    &  -1.721 $\pm$ 0.003 &   0.032 $\pm$ 0.005 &  0.89 $\pm$ 0.02 &   8.39$_{-1.07}^{+1.19}$   \\
    \bottomrule
    \end{tabular} \\
\caption{Same as Table \ref{tab:yu_k}, except for a subsample of stars from the APOKASC-2 \citep{art:pinsonneault+2018} sample.}
\label{tab:apo_k}
\end{table*}

\begin{table*}
    \begin{tabular}{rrrrr|rrrr}
    \toprule
    {} & \multicolumn{4}{|c|}{No Correction} & \multicolumn{4}{|c|}{Clump Corrected} \\
    \cmidrule(lr){2-5}\cmidrule(lr){6-9}
    \multicolumn{1}{c}{$\Delta T_{\rm eff}\ (K)$} & \multicolumn{1}{c}{$\mu_{\rm RC}\ (\rm mag)$} & \multicolumn{1}{c}{$\sigma_{\rm RC}\ (\rm mag)$}  & \multicolumn{1}{c}{$Q$} &  \multicolumn{1}{c}{$\sigma_{\rm o}\ (\sigma_{\rm RC})$} & \multicolumn{1}{c}{$\mu_{\rm RC}\ (\rm mag)$} & \multicolumn{1}{c}{$\sigma_{\rm RC}\ (\rm mag)$}  & \multicolumn{1}{c}{$Q$} &  \multicolumn{1}{c}{$\sigma_{\rm o}\ (\sigma_{\rm RC})$} \\
    \midrule
        \textbf{-50.0} &   0.35 $\pm$ 0.003 &  0.181 $\pm$ 0.004 &  0.98 $\pm$ 0.01 &  2.73$_{-0.42}^{+0.58}$  &   0.34 $\pm$ 0.003 &   0.193 $\pm$ 0.004 &  0.99 $\pm$ 0.01 &   2.77$_{-0.46}^{+0.66}$ \\
        \textbf{-40.0} &  0.336 $\pm$ 0.003 &  0.181 $\pm$ 0.004 &  0.98 $\pm$ 0.01 &  2.73$_{- 0.4}^{+0.56}$  &  0.325 $\pm$ 0.003 &   0.192 $\pm$ 0.004 &  0.99 $\pm$ 0.01 &   2.78$_{-0.45}^{+0.66}$  \\
        \textbf{-30.0} &  0.323 $\pm$ 0.003 &   0.18 $\pm$ 0.004 &  0.98 $\pm$ 0.01 &  2.72$_{- 0.4}^{+0.54}$  &  0.311 $\pm$ 0.003 &    0.19 $\pm$ 0.004 &  0.99 $\pm$ 0.01 &   2.79$_{-0.43}^{+0.58}$ \\
        \textbf{-20.0} &  0.309 $\pm$ 0.003 &  0.179 $\pm$ 0.004 &  0.98 $\pm$ 0.01 &  2.72$_{- 0.4}^{+0.54}$  &  0.297 $\pm$ 0.003 &   0.188 $\pm$ 0.004 &  0.98 $\pm$ 0.01 &   2.72$_{-0.41}^{+0.58}$  \\
        \textbf{-10.0} &  0.295 $\pm$ 0.003 &  0.178 $\pm$ 0.004 &  0.98 $\pm$ 0.01 &  2.69$_{-0.38}^{+0.53}$  &  0.282 $\pm$ 0.003 &   0.187 $\pm$ 0.004 &  0.98 $\pm$ 0.01 &   2.71$_{- 0.4}^{+0.58}$ \\
        \textbf{0.0  } &  0.282 $\pm$ 0.003 &  0.177 $\pm$ 0.004 &  0.98 $\pm$ 0.01 &  2.68$_{-0.38}^{+0.52}$  &  0.268 $\pm$ 0.003 &   0.187 $\pm$ 0.004 &  0.98 $\pm$ 0.01 &   2.73$_{-0.42}^{+0.58}$  \\
        \textbf{10.0 } &  0.268 $\pm$ 0.003 &  0.177 $\pm$ 0.004 &  0.98 $\pm$ 0.01 &  2.71$_{-0.39}^{+0.53}$  &  0.254 $\pm$ 0.003 &   0.185 $\pm$ 0.004 &  0.98 $\pm$ 0.01 &    2.7$_{-0.41}^{+0.58}$ \\
        \textbf{20.0 } &  0.255 $\pm$ 0.003 &  0.176 $\pm$ 0.004 &  0.98 $\pm$ 0.01 &   2.7$_{-0.37}^{+0.51}$  &   0.24 $\pm$ 0.003 &   0.184 $\pm$ 0.004 &  0.98 $\pm$ 0.01 &   2.71$_{- 0.4}^{+0.56}$  \\
        \textbf{30.0 } &  0.241 $\pm$ 0.003 &  0.175 $\pm$ 0.004 &  0.98 $\pm$ 0.01 &  2.68$_{-0.36}^{+0.51}$  &  0.226 $\pm$ 0.003 &   0.183 $\pm$ 0.004 &  0.98 $\pm$ 0.01 &    2.7$_{- 0.4}^{+0.55}$ \\
        \textbf{40.0 } &  0.228 $\pm$ 0.003 &  0.174 $\pm$ 0.004 &  0.98 $\pm$ 0.01 &  2.67$_{-0.36}^{+0.48}$  &  0.213 $\pm$ 0.003 &   0.182 $\pm$ 0.004 &  0.98 $\pm$ 0.01 &    2.7$_{- 0.4}^{+0.52}$  \\
        \textbf{50.0 } &  0.215 $\pm$ 0.003 &  0.173 $\pm$ 0.004 &  0.98 $\pm$ 0.01 &  2.68$_{-0.36}^{+0.48}$  &  0.199 $\pm$ 0.003 &   0.181 $\pm$ 0.004 &  0.98 $\pm$ 0.01 &   2.69$_{-0.38}^{+0.53}$ \\
    \bottomrule
    \end{tabular} \\
\caption{Same as Table \ref{tab:yu_k}, except for the \gaia $G$ band, for \nstars stars from the \citetalias{art:yu+2018} sample.}
\label{tab:yu_g}
\end{table*}

\begin{table*}
    \begin{tabular}{rrrrr|rrrr}
    \toprule
    {} & \multicolumn{4}{|c|}{No Correction} & \multicolumn{4}{|c|}{Clump Corrected} \\
    \cmidrule(lr){2-5}\cmidrule(lr){6-9}
    \multicolumn{1}{c}{$\Delta T_{\rm eff}\ (K)$} & \multicolumn{1}{c}{$\mu_{\rm RC}\ (\rm mag)$} & \multicolumn{1}{c}{$\sigma_{\rm RC}\ (\rm mag)$}  & \multicolumn{1}{c}{$Q$} &  \multicolumn{1}{c}{$\sigma_{\rm o}\ (\sigma_{\rm RC})$} & \multicolumn{1}{c}{$\mu_{\rm RC}\ (\rm mag)$} & \multicolumn{1}{c}{$\sigma_{\rm RC}\ (\rm mag)$}  & \multicolumn{1}{c}{$Q$} &  \multicolumn{1}{c}{$\sigma_{\rm o}\ (\sigma_{\rm RC})$} \\
    \midrule
        \textbf{-50.0} &   0.53 $\pm$ 0.004 &  0.118 $\pm$ 0.006 &  0.96$_{-0.03}^{+0.02}$ &  3.19$_{-0.46}^{+0.65}$    &  0.526 $\pm$ 0.004 &   0.128 $\pm$ 0.005 &  0.97$_{-0.02}^{+0.01}$ &   3.29$_{-0.55}^{+0.81}$ \\
        \textbf{-40.0} &  0.516 $\pm$ 0.004 &  0.117 $\pm$ 0.005 &  0.96$_{-0.03}^{+0.02}$ &  3.18$_{-0.46}^{+0.65}$    &   0.51 $\pm$ 0.004 &   0.127 $\pm$ 0.005 &  0.97$_{-0.02}^{+0.01}$ &   3.31$_{-0.55}^{+ 0.8}$ \\
        \textbf{-30.0} &  0.501 $\pm$ 0.004 &  0.116 $\pm$ 0.006 &  0.96$_{-0.03}^{+0.02}$ &  3.19$_{-0.45}^{+0.62}$    &  0.495 $\pm$ 0.004 &   0.127 $\pm$ 0.005 &  0.97$_{-0.02}^{+0.01}$ &   3.29$_{-0.54}^{+0.76}$ \\
        \textbf{-20.0} &  0.486 $\pm$ 0.004 &  0.116 $\pm$ 0.006 &  0.96$_{-0.03}^{+0.02}$ &  3.19$_{-0.47}^{+0.65}$    &  0.479 $\pm$ 0.004 &   0.126 $\pm$ 0.005 &  0.97$_{-0.02}^{+0.01}$ &   3.28$_{-0.54}^{+0.77}$ \\
        \textbf{-10.0} &  0.472 $\pm$ 0.004 &  0.115 $\pm$ 0.006 &  0.96$_{-0.03}^{+0.02}$ &  3.19$_{-0.45}^{+0.63}$    &  0.464 $\pm$ 0.004 &   0.126 $\pm$ 0.005 &  0.97$_{-0.02}^{+0.01}$ &   3.29$_{-0.55}^{+0.74}$ \\
        \textbf{0.0  } &  0.457 $\pm$ 0.004 &  0.114 $\pm$ 0.006 &  0.96$_{-0.03}^{+0.02}$ &   3.2$_{-0.45}^{+0.64}$    &  0.449 $\pm$ 0.004 &   0.125 $\pm$ 0.005 &  0.97$_{-0.02}^{+0.01}$ &   3.27$_{-0.53}^{+0.79}$ \\
        \textbf{10.0 } &  0.443 $\pm$ 0.004 &  0.113 $\pm$ 0.006 &  0.95$_{-0.03}^{+0.02}$ &  3.17$_{-0.44}^{+0.63}$    &  0.434 $\pm$ 0.004 &   0.124 $\pm$ 0.005 &  0.97$_{-0.02}^{+0.01}$ &   3.25$_{-0.52}^{+ 0.8}$ \\
        \textbf{20.0 } &  0.429 $\pm$ 0.004 &  0.113 $\pm$ 0.006 &  0.96$_{-0.03}^{+0.02}$ &  3.21$_{-0.44}^{+0.62}$    &  0.419 $\pm$ 0.004 &   0.124 $\pm$ 0.005 &  0.97$_{-0.02}^{+0.01}$ &   3.25$_{-0.53}^{+0.73}$ \\
        \textbf{30.0 } &  0.414 $\pm$ 0.004 &  0.112 $\pm$ 0.006 &  0.95$_{-0.03}^{+0.02}$ &  3.18$_{-0.43}^{+0.61}$    &  0.404 $\pm$ 0.004 &   0.123 $\pm$ 0.005 &  0.97$_{-0.02}^{+0.01}$ &   3.25$_{-0.54}^{+0.72}$ \\
        \textbf{40.0 } &    0.4 $\pm$ 0.004 &  0.112 $\pm$ 0.006 &  0.95$_{-0.03}^{+0.02}$ &  3.19$_{-0.44}^{+0.59}$    &  0.389 $\pm$ 0.004 &   0.122 $\pm$ 0.005 &  0.97$_{-0.02}^{+0.02}$ &   3.24$_{- 0.5}^{+0.72}$ \\
        \textbf{50.0 } &  0.386 $\pm$ 0.004 &  0.111 $\pm$ 0.006 &  0.95$_{-0.03}^{+0.02}$ &   3.2$_{-0.43}^{+0.57}$    &  0.375 $\pm$ 0.004 &   0.122 $\pm$ 0.006 &  0.97$_{-0.02}^{+0.01}$ &   3.25$_{-0.51}^{+0.71}$ \\
    \bottomrule
    \end{tabular} \\
\caption{Same as Table \ref{tab:yu_k}, except for the \gaia $G$ band, for a subsample of stars from the APOKASC-2 \citep{art:pinsonneault+2018} sample.}
\label{tab:apo_g}
\end{table*}

\begin{table*}
    \begin{tabular}{rrrrrrr}
    \toprule
    \multicolumn{1}{c}{$\Delta T_{\rm eff}\ (K)$} & \multicolumn{1}{c}{$\mu_{\rm RC}\ (\rm mag)$} & \multicolumn{1}{c}{$\sigma_{\rm RC}\ (\rm mag)$}  & \multicolumn{1}{c}{$Q$} &  \multicolumn{1}{c}{$\sigma_{\rm o}\ (\sigma_{\rm RC})$} & \multicolumn{1}{c}{$L\ (\rm pc)$} & \multicolumn{1}{c}{$\varpi_{\rm zp}\ (\mu \rm as)$} \\
    \midrule
        \textbf{-50.0} &  -1.71 $\pm$ 0.002 &  0.041 $\pm$ 0.003 &  0.58$_{-0.05}^{+0.05}$ &  5.49$_{-0.47}^{+0.52}$ &  908.63$_{-15.89}^{+16.55}$ & -24.09$_{-12.76}^{+12.84}$ \\
        \textbf{0.0  } & -1.737 $\pm$ 0.002 &   0.04 $\pm$ 0.003 &  0.55$_{-0.03}^{+0.05}$ &  5.61$_{-0.47}^{+ 0.5}$ &  920.12$_{-16.61}^{+17.18}$ &  -19.5$_{-12.46}^{+ 12.4}$ \\
        \textbf{50.0 } & -1.764 $\pm$ 0.002 &  0.041 $\pm$ 0.004 &  0.53$_{-0.02}^{+0.04}$ &   5.5$_{-0.48}^{+ 0.5}$ &  930.95$_{-16.83}^{+18.07}$ & -14.81$_{-12.98}^{+12.57}$ \\
    \bottomrule
    \end{tabular}
\caption{Medians of the posterior distributions for hyperparameters of our \gaia model, for the 2MASS $K$ band, for a randomly selected subsample of 1000 stars from the \citetalias{art:yu+2018} sample. Uncertainties are taken as the 1$\sigma$ credible intervals, and are listed as a single value for cases where the posterior was approximately Gaussian. Priors were imposed on \murc and \sigrc corresponding to the results for these values using seismic Clump Corrected data in Table \ref{tab:yu_k}, for the temperature shifts shown in the $\Delta\teff$ column. $L$ is the length scale of the exponentially decaying space density prior on distance \citep{art:bailer-jones+2018}, and \oozp is the parallax zero-point offset. All other symbols are the same as for Table \ref{tab:yu_k}.}
\label{tab:gaia_yu_k}
\end{table*}

\begin{table*}
    \begin{tabular}{rrrrrrr}
    \toprule
    \multicolumn{1}{c}{$\Delta T_{\rm eff}\ (K)$} & \multicolumn{1}{c}{$\mu_{\rm RC}\ (\rm mag)$} & \multicolumn{1}{c}{$\sigma_{\rm RC}\ (\rm mag)$}  & \multicolumn{1}{c}{$Q$} &  \multicolumn{1}{c}{$\sigma_{\rm o}\ (\sigma_{\rm RC})$} & \multicolumn{1}{c}{$L\ (\rm pc)$} & \multicolumn{1}{c}{$\varpi_{\rm zp}\ (\mu \rm as)$} \\
    \midrule
        \textbf{-50.0} & -1.661 $\pm$ 0.003 &   0.04 $\pm$ 0.003 &   0.6 $\pm$ 0.05 &  5.76$_{- 0.5}^{+0.55}$   &     888$_{-15.78}^{+16.38}$  & -33.53$_{-12.97}^{+12.93}$\\
        \textbf{0.0  } & -1.689 $\pm$ 0.003 &   0.04 $\pm$ 0.004 &  0.59 $\pm$ 0.05 &  5.66$_{-0.51}^{+0.53}$   &  899.36$_{-16.04}^{+16.72}$  & -28.33$_{-12.92}^{+12.96}$ \\
        \textbf{50.0 } & -1.715 $\pm$ 0.003 &  0.041 $\pm$ 0.004 &  0.57 $\pm$ 0.05 &  5.51$_{-0.49}^{+0.55}$   &  910.68$_{- 16.4}^{+16.83}$  & -23.47$_{-13.13}^{+13.25}$\\
    \bottomrule
    \end{tabular}
\caption{Same as Table \ref{tab:gaia_yu_k}, except for priors imposed on \murc and \sigrc corresponding to the results for these values using seismic Clump Corrected data in Table \ref{tab:apo_k}, for the temperature shifts shown in the $\Delta\teff$ column.}
\label{tab:gaia_apo_k}
\end{table*}

\begin{table*}
    \begin{tabular}{rrrrrrr}
    \toprule
    \multicolumn{1}{c}{$\Delta T_{\rm eff}\ (K)$} & \multicolumn{1}{c}{$\mu_{\rm RC}\ (\rm mag)$} & \multicolumn{1}{c}{$\sigma_{\rm RC}\ (\rm mag)$}  & \multicolumn{1}{c}{$Q$} &  \multicolumn{1}{c}{$\sigma_{\rm o}\ (\sigma_{\rm RC})$} & \multicolumn{1}{c}{$L\ (\rm pc)$} & \multicolumn{1}{c}{$\varpi_{\rm zp}\ (\mu \rm as)$} \\
    \midrule
        \textbf{-50.0} &  0.346 $\pm$ 0.003 &   0.19 $\pm$ 0.003 &  0.97$_{-0.02}^{+0.01} $ &   3.1$_{- 0.7}^{+0.79}$&   948.41$_{-17.96}^{+18.15}$ &  -9.96$_{-13.18}^{+ 13.1}$\\
        \textbf{0.0  } &  0.277 $\pm$ 0.003 &  0.188 $\pm$ 0.004 &  0.95$_{-0.05}^{+0.02} $ &  2.64$_{-0.63}^{+0.77}$&    978.9$_{-17.46}^{+18.35}$ &   1.14$_{-12.81}^{+ 12.8}$\\
        \textbf{50.0 } &  0.209 $\pm$ 0.003 &  0.184 $\pm$ 0.004 &  0.74$_{-0.13}^{+0.12} $ &  1.71$_{- 0.2}^{+0.36}$&  1008.77$_{-18.48}^{+18.85}$ &  10.76$_{-13.21}^{+13.13}$\\
    \bottomrule
    \end{tabular}
\caption{Same as Table \ref{tab:gaia_yu_k}, except for the \gaia $G$ band, with priors imposed on \murc and \sigrc corresponding to the results for these values using seismic Clump Corrected data in Table \ref{tab:yu_g}, for the temperature shifts shown in the $\Delta\teff$ column.}
\label{tab:gaia_yu_g}
\end{table*}

\begin{table*}
    \begin{tabular}{rrrrrrr}
    \toprule
    \multicolumn{1}{c}{$\Delta T_{\rm eff}\ (K)$} & \multicolumn{1}{c}{$\mu_{\rm RC}\ (\rm mag)$} & \multicolumn{1}{c}{$\sigma_{\rm RC}\ (\rm mag)$}  & \multicolumn{1}{c}{$Q$} &  \multicolumn{1}{c}{$\sigma_{\rm o}\ (\sigma_{\rm RC})$} & \multicolumn{1}{c}{$L\ (\rm pc)$} & \multicolumn{1}{c}{$\varpi_{\rm zp}\ (\mu \rm as)$} \\
    \midrule
        \textbf{-50.0} &  0.527 $\pm$ 0.004 &   0.13 $\pm$ 0.005 &  0.82$_{-0.07}^{+0.05}$ &  2.53$_{- 0.3}^{+0.36}$ &  874.12$_{- 16.1}^{+16.56}$  & -39.02$_{-13.16}^{+12.98}$\\
        \textbf{0.0  } &  0.455 $\pm$ 0.004 &  0.129 $\pm$ 0.005 &  0.79$_{-0.08}^{+0.07}$ &  2.42$_{-0.29}^{+0.36}$ &  903.23$_{-16.68}^{+ 16.8}$  & -26.84$_{-12.97}^{+ 13.1}$ \\
        \textbf{50.0 } &  0.385 $\pm$ 0.004 &  0.127 $\pm$ 0.005 &  0.68$_{- 0.1}^{+0.09}$ &  2.22$_{-0.22}^{+0.27}$ &  931.92$_{- 17  }^{+17.53}$  & -14.94$_{-13.04}^{+12.58}$\\
    \bottomrule
    \end{tabular}
\caption{Same as Table \ref{tab:gaia_yu_k}, except for the \gaia $G$ band, with priors imposed on \murc and \sigrc corresponding to the results for these values using seismic Clump Corrected data in Table \ref{tab:apo_g} , for the temperature shifts shown in the $\Delta\teff$ column.} 
\label{tab:gaia_apo_g}
\end{table*}

\begin{table*}
    \begin{tabular}{llrrrrrr}
    \toprule
    \multicolumn{1}{c}{Source} &  \multicolumn{1}{c}{$\varpi_{\rm zp}$ prior $(\mu \rm as)$} & \multicolumn{1}{c}{$\mu_{\rm RC}\ (\rm mag)$} & \multicolumn{1}{c}{$\sigma_{\rm RC}\ (\rm mag)$}  & \multicolumn{1}{c}{$Q$} &  \multicolumn{1}{c}{$\sigma_{\rm o}\ (\sigma_{\rm RC})$} & \multicolumn{1}{c}{$L\ (\rm pc)$} & \multicolumn{1}{c}{$\varpi_{\rm zp}\ (\mu \rm as)$} \\
    \midrule
        \textbf{Lindegren+ 18 } &                   $\mathcal{N}(-29.0 , 1.0) $  & -1.638 $\pm$ 0.017 &  0.075$_{-0.015}^{+0.016}$ &  0.78$_{-0.11}^{+0.09}  $   &  3.28$_{-0.56}^{+0.64}$ &  888.56$_{-24.36}^{+25.41} $ & -29.07$_{- 0.99 }^{+     1 }$\\
        \textbf{Zinn+ 18      } &                   $\mathcal{N}(-52.8 , 3.4) $  & -1.631 $\pm$ 0.017 &  0.074$_{-0.015}^{+0.016}$ &  0.77$_{- 0.1}^{+0.09}  $   &   3.3$_{-0.57}^{+0.65}$ &  885.76$_{-23.73}^{+ 24.4} $ & -51.92$_{- 3.21 }^{+  3.21 }$\\
        \textbf{Riess+ 18     } &                   $\mathcal{N}(-46.0 , 13.0)$  & -1.634 $\pm$ 0.017 &  0.076$_{-0.015}^{+0.017}$ &  0.78$_{-0.11}^{+0.09}  $   &  3.26$_{-0.57}^{+0.64}$ &  886.59$_{-24.12}^{+   25} $ & -42.22$_{- 9.16 }^{+  9.33 }$\\
        \textbf{Sahlholdt \& Silva Aguirre18 } &    $\mathcal{N}(-35.0 , 16.0)$  & -1.634 $\pm$ 0.017 &  0.073$_{-0.015}^{+0.016}$ &  0.77$_{-0.11}^{+0.09}  $   &  3.33$_{-0.58}^{+0.69}$ &  887.37$_{-23.89}^{+24.06} $ &    -37$_{-10.42 }^{+ 10.17 }$\\
        \textbf{Stassun \& Torres 18   } &          $\mathcal{N}(-82.0 , 33.0)$  & -1.632 $\pm$ 0.017 &  0.072$_{-0.016}^{+0.017}$ &  0.76$_{-0.11}^{+0.09}  $   &  3.36$_{-0.59}^{+0.64}$ &  885.77$_{-23.01}^{+24.33} $ & -44.55$_{-12.59 }^{+ 12.62 }$\\
        \textit{Hawkins+ 17 } &                     $\mathcal{N}(0.0 , 1.0)   $  & -1.648 $\pm$ 0.018 &  0.075$_{-0.015}^{+0.017}$ &  0.78$_{-0.11}^{+0.09}  $   &  3.31$_{-0.57}^{+0.64}$ &  893.39$_{-24   }^{+24.6 } $ &  -0.22$_{-  1.01}^{+   0.99}$ \\
        \textit{Uninformed} &                       $\mathcal{N}(0.0 , 1000.0)$  & -1.634 $\pm$ 0.018 &  0.074$_{-0.015}^{+0.017}$ &  0.77$_{-0.11}^{+0.09}  $   &   3.3$_{-0.58}^{+0.64}$ &  887.27$_{-23.82}^{+24.12} $ & -38.38$_{-13.54 }^{+ 13.83 }$\\
    \bottomrule
    \end{tabular}
\caption{Medians on the posterior distributions for hyperparameters on our \gaia model, for the 2MASS $K$ band, for a randomly selected subsample of 1000 stars from the \citetalias{art:yu+2018} sample. Uncertainties are taken as \nnew{the 1$\sigma$ credible intervals}, and are listed as single values for cases where the posterior was approximately Gaussian. Highly informative priors, shown in the `\oozp prior' column, were imposed on \oozp corresponding to estimates for this parameter from the literature, listed in bold print in the Source column. Additionally, we apply a custom prior to place \oozp near zero in order to recreate conditions similar to the \citetalias{art:hawkins+2017} work, and an extremely broad prior on \oozp in order to find a value given no strong constraints on neither \oozp, \murc or \sigrc. $\mathcal{N}(\mu, \sigma)$ indicates a normal distribution with mean $\mu$ and standard deviation $\sigma$.}
\label{tab:parallax_k}
\end{table*}

\begin{table*}
    \begin{tabular}{llrrrrrr}
    \toprule
    \multicolumn{1}{c}{Source} &  \multicolumn{1}{c}{$\varpi_{\rm zp}$ prior $(\mu \rm as)$} & \multicolumn{1}{c}{$\mu_{\rm RC}\ (\rm mag)$} & \multicolumn{1}{c}{$\sigma_{\rm RC}\ (\rm mag)$}  & \multicolumn{1}{c}{$Q$} &  \multicolumn{1}{c}{$\sigma_{\rm o}\ (\sigma_{\rm RC})$} & \multicolumn{1}{c}{$L\ (\rm pc)$} & \multicolumn{1}{c}{$\varpi_{\rm zp}\ (\mu \rm as)$} \\
    \midrule
      \textbf{Lindegren+ 18 }                     &  $\mathcal{N}(-29.0 , 1.0)  $ &  0.542 $\pm$ 0.016 &  0.138$_{-0.018}^{+0.014}$ &  0.86$_{-0.12}^{+0.07}$ &  2.61$_{-0.34}^{+0.48}$ &   868.2$_{-17.09}^{+17.41}$ & -29.06$_{- 1.01}^{+ 0.98}$ \\
      \textbf{Zinn+ 18      }                     &  $\mathcal{N}(-52.8 , 3.4)  $ &  0.548 $\pm$ 0.016 &  0.139$_{-0.018}^{+0.014}$ &  0.86$_{-0.12}^{+0.07}$ &  2.62$_{-0.35}^{+0.49}$ &  865.44$_{-17.15}^{+16.95}$ & -52.18$_{- 3.31}^{+ 3.27}$ \\
      \textbf{Riess+ 18     }                     &  $\mathcal{N}(-46.0 , 13.0) $ &  0.545 $\pm$ 0.016 &   0.14$_{-0.017}^{+0.013}$ &  0.87$_{-0.11}^{+0.07}$ &  2.62$_{-0.34}^{+0.48}$ &  867.13$_{-17.55}^{+17.23}$ & -44.23$_{- 9.32}^{+ 9.06}$ \\
      \textbf{Sahlholdt \& Silva Aguirre 18 }    &  $\mathcal{N}(-35.0 , 16.0) $  &  0.545 $\pm$ 0.016 &  0.136$_{-0.021}^{+0.015}$ &  0.85$_{-0.14}^{+0.08}$ &  2.62$_{-0.34}^{+0.47}$ &  867.15$_{-17.05}^{+ 17.3}$ & -39.29$_{-10.27}^{+ 9.86}$ \\
      \textbf{Stassun \& Torres 18   }           &  $\mathcal{N}(-82.0 , 33.0) $  &  0.546 $\pm$ 0.017 &  0.138$_{-0.018}^{+0.014}$ &  0.86$_{-0.12}^{+0.07}$ &  2.61$_{-0.33}^{+0.46}$ &  866.11$_{-17.02}^{+17.76}$ & -47.86$_{-12.51}^{+12.18}$ \\
      \textit{Hawkins+ 17   }                     &  $\mathcal{N}(0.0 , 1.0)    $ &  0.534 $\pm$ 0.015 &   0.14$_{-0.018}^{+0.013}$ &  0.87$_{-0.12}^{+0.06}$ &  2.64$_{-0.35}^{+ 0.5}$ &  872.01$_{-17.38}^{+ 17.8}$ &  -0.23$_{- 1.01}^{+    1}$ \\
      \textit{Uninformed}                     &  $\mathcal{N}(0.0 , 1000.0) $     &  0.546 $\pm$ 0.016 &  0.139$_{-0.019}^{+0.013}$ &  0.87$_{-0.13}^{+0.07}$ &  2.62$_{-0.34}^{+0.49}$ &  866.26$_{-16.86}^{+17.53}$ & -42.66$_{-13.14}^{+13.48}$ \\
    \bottomrule
    \end{tabular}
\caption{Same as Table \ref{tab:parallax_k}, except for the \gaia $G$ band.}
\label{tab:parallax_g}
\end{table*}

\begin{figure*}
    \centering
    \includegraphics[width=\textwidth]{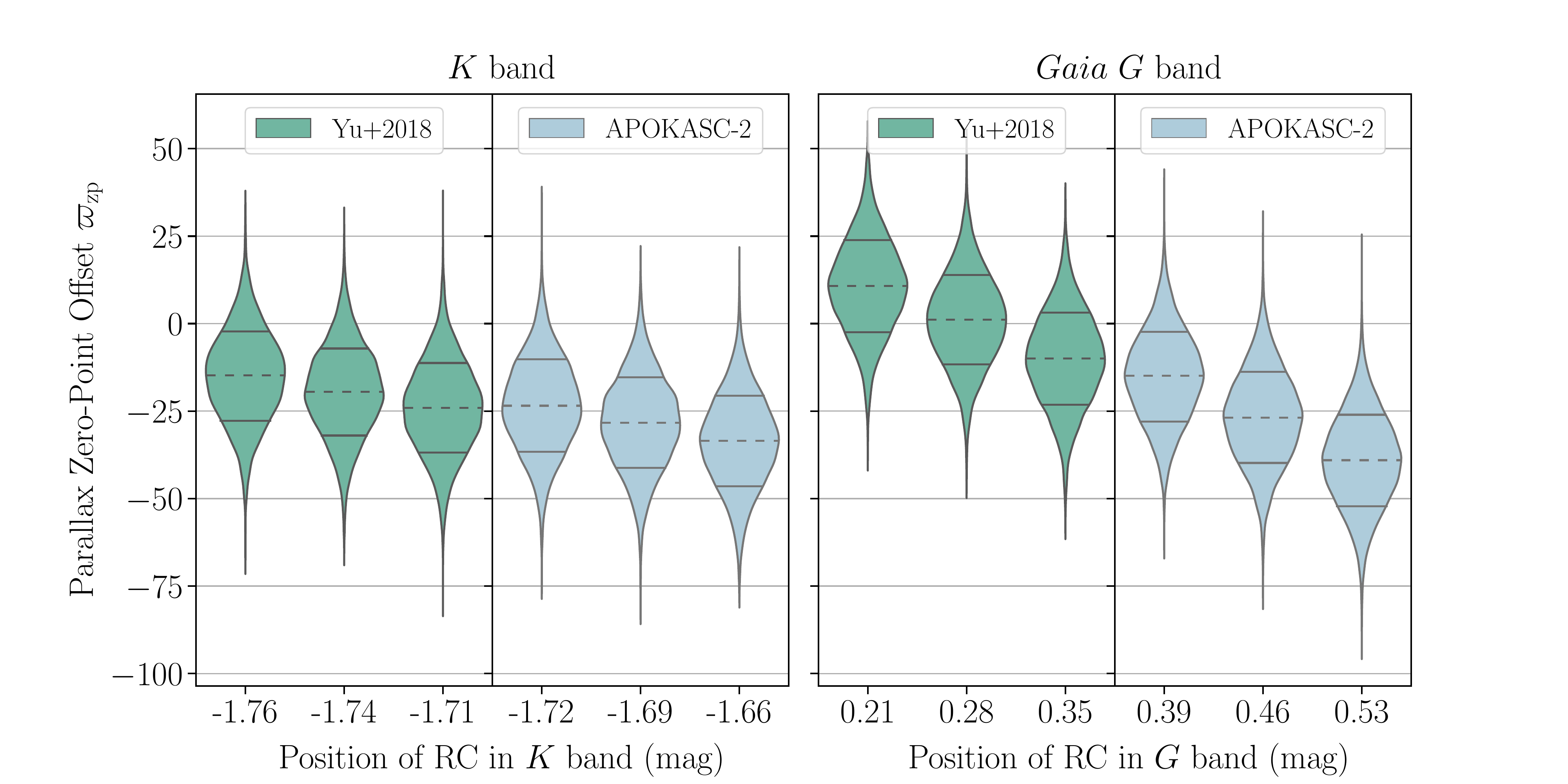}
    \caption{The posterior distributions on the parallax zero-point offset \oozp, as a function of the absolute magnitude of the RC used to calibrate this value, for 1000 randomly selected stars across the \kepler field. The RC magnitudes on the x-axis correspond to those obtained from seismology for perturbations to the temperature values $\Delta T_{\rm eff}$ of $-50$, $0$, and $+50\ \rm K$, from runs on our full sample \protect\citep{art:yu+2018} and the APOKASC-2 sample \protect\citep{art:pinsonneault+2018}. The dashed horizontal lines indicate the median on the posteriors, and the solid horizontal lines represent \nnew{the 1$\sigma$ credible intervals}. The posteriors' magnitudes along the x-axis are indicative of power with arbitrary units, whereas their shape along the y-axis indicates the spread in the posterior result, and is reflected across the x-axis.}
    \label{fig:gaia_posteriors}    
\end{figure*}

\subsection{Results from \gaia}
Given our results from asteroseismology, we wish to determine the parallax zero-point offset, \oozp, that recovers our values of the absolute magnitude and spread of the RC. Since $\mu_{\rm RC}$ and $\sigma_{\rm RC}$ represent astrophysical observables that should be consistent across both data sets, we used a description of the posterior distributions from these parameters from our seismic model as a highly informative prior in our \gaia model. This yields the parallax offset required to recover the same magnitude and spread of the RC found using seismology. We passed in the seismic posteriors for $\Delta T_{\rm eff}$ being $-50$, $0$, and $+50\ \rm K$ from our runs on our full sample and the reduced APOKASC-2 sample, and thus ran our model for 6 different RC magnitudes \& spreads in each band. Additionally, we used the median values of each latent parameter $M_i$ from the application of our seismic model to our full sample, along with distance estimates by \cite{art:bailer-jones+2018} and observed extinctions from \cite{art:green+2018}, as initial guesses in our \gaia model for computational efficiency. No other values were changed on each run.

Following the relation presented in equation \ref{eq:covariance} \citep{art:lindegren+2018} we treated our parallax uncertainties as correlated as a function of position on the sky across the entire \kepler field, similarly to previous work by \cite{art:zinn+2018}. While the model equation presented by \cite{art:lindegren+2018} describes the covariance well for a wide range of separations, individual covariances oscillate around the model at separations below 1 deg, and the model no longer holds at all for separations below 0.125 deg. To ensure that our treatment of the parallax covariances was sensible, we ran our \gaia model on a reduced sample of 1000 stars, randomly selected from across the entire \kepler field to ensure sparsity. This reduced sample contained no angular separations in the range $< 0.125\ \rm deg$\footnote{The data were shuffled using the \texttt{sklearn.utils.shuffle} function with a random seed of 24601.}.

In Tables \ref{tab:gaia_yu_k}, \ref{tab:gaia_apo_k}, \ref{tab:gaia_yu_g} and \ref{tab:gaia_apo_g} we present the medians on the posterior distributions of our hyperparameters for our \gaia model, given RC-corrected seismic positions and spreads for the RC at different temperature offsets $\Delta T_{\rm eff}$ for both the \citetalias{art:yu+2018} and APOKASC-2 samples. In Figure \ref{fig:gaia_posteriors}, we present the posterior distributions of \oozp given the 6 values for the position of the RC used each in the $K$ and $G$ bands.\\

In order to probe the impact of literature values for \oozp on an inference of our RC parameters, we reran our \gaia model for the $K$ and \gaia $G$ bands with a strongly informative prior on \oozp (see equation \ref{eq:zeropointpriors}). We did this for the same reduced sample of 1000 stars from our \citetalias{art:yu+2018} sample. For all these runs, we applied the same priors used for \murc and \sigrc as in the asteroseismic runs (see equation \ref{eq:asteropriors}). We used the parallax zero-point offsets reported by \cite{art:lindegren+2018} ($-29$ \muas, with an assumed uncertainty of $1$ \muas), \cite{art:zinn+2018} ($-52.8$ \muas with a total uncertainty of $3.4$ \muas), \cite{art:riess+2018} ($-46 \pm 13$ \muas), \cite{art:sahlholdt+silvaaguirre2018} ($-35 \pm 16$ \muas) and \cite{art:stassun+torres2018} ($-82 \pm 33$ \muas). \up{Note that for the purpose of calibration not all these zero-point offsets would be applicable to our sample due to differences in colour, magnitude, and position. We instead used them as representative of \oozp in the literature to study their impact on our inferences only.} In addition, we also ran with a prior of $0 \pm 1$ \muas in an attempt to recreate the \citetalias{art:hawkins+2017} work (albeit accounting for parallax covariances), as well as a single run with no strongly informative priors on \oozp, \murc or \sigrc, \up{thus finding our own measure of the zero-point offset.}

\begin{figure*}
    \centering
    \includegraphics[width=\textwidth]{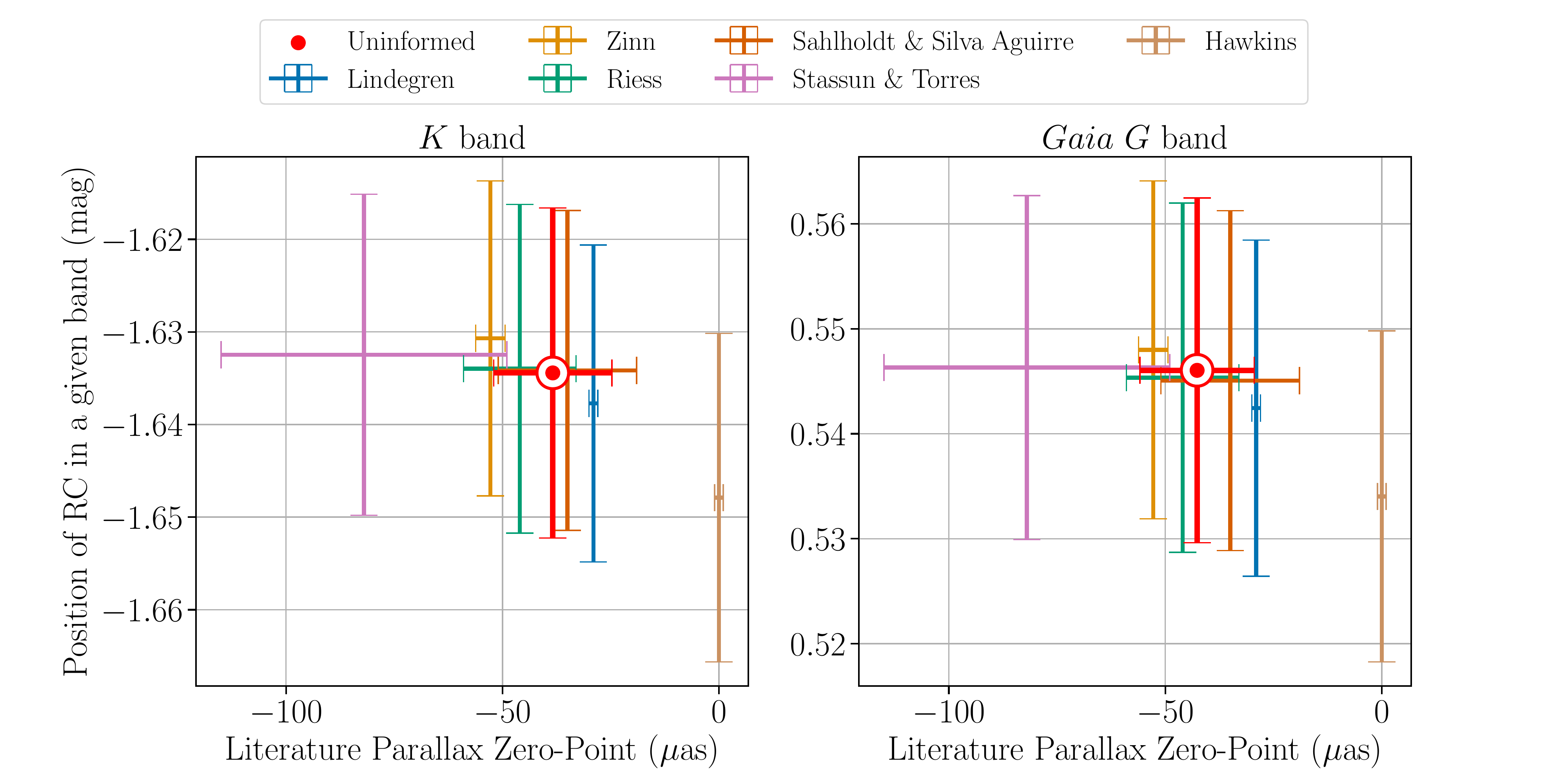}
    \caption{The \nnew{the 1$\sigma$ credible intervals} for the posterior distributions on \murc, as a function of the value for \oozp used as an informative prior on \oozp, for 1000 randomly selected stars across the \textit{Kepler} field in both the 2MASS $K$ and \gaia $G$ bands. The errorbars on the x-axis correspond to the formal uncertainties for literature values, or are otherwise specified in the text. the `uninformed' value corresponds to a run of our \gaia model with no strong constrains on \oozp, and in this case the x-axis errobars correspond go \nnew{the 1$\sigma$ credible intervals} on the inferred value for \oozp.}
    \label{fig:parallax_values}    
\end{figure*}

In Tables \ref{tab:parallax_k} \& \ref{tab:parallax_g} we present the medians and $1\sigma$ credible intervals on the posterior distributions for the hyperparameters of our \gaia model given the conditions stated above, as well as naming the source of the used parallax zero-point offset, and an expression of the prior applied to \oozp. Note that \nnew{the} inferred value of \oozp may differ significantly within the uncertainties on any of the literature values used. In Figure \ref{fig:parallax_values} we present the medians and \nnew{$1\sigma$ credible intervals} on the posterior distributions for \murc given our chosen values for \oozp, with the result from the `uninformed' run shown with bold red error bars.

\section{Discussion}\label{sec:discussion}
\subsection{Luminosity of the Red Clump}
\nnew{Figures \ref{fig:yu_posteriors} and \ref{fig:apokasc2_posteriors} show the posteriors on the inferred absolute magnitude of the RC, \murc, for the $K$ and \gaia $G$ bands given changes to effective temperature and corrections to the scaling relations.} There is a clear relation between the overall offset in \teff and the inferred magnitude of the RC, where a change of about $20\ \rm{K}$ results in a difference of more than $1\sigma$. The overall relation between the clump magnitude and temperature is expected, given the large impact of temperature on the calculations for absolute magnitude; luminosity calculated via the seismic scaling relations scales with temperature to a power of 4.5, and bolometric corrections calculated through the \cite{art:casagrande+vandenberg2018} method rely on both \teff and \logg, which is calculated using \teff. \new{The small uncertainties on \murc and \sigrc indicate the ability of hierarchical models to leverage a large number of individual uncertainties to fit to a population model, given that the uncertainties on our data for \teff are well above the shifts in temperature we are applying.}

We also see that the scaling relation corrections appear to be degenerate with a small temperature offset. A change of $\sim 20\ \rm{K}$ to the temperatures provides a similar clump magnitude as when applying a correction to the scaling relations. At higher temperatures, the difference in the magnitude of the RC between corrected and uncorrected scaling relations increases. This shows that the \teff values have a significant impact on the \fdnu obtained through the \cite{art:sharma+stello2016} method, even at relatively small \teff shifts. 

The values for \murc in both bands are fainter for the subset of stars using APOKASC-2 temperatures than those using temperatures from \cite{art:mathur+2017}. This reflects the relation we already saw between \teff and \murc for the \citetalias{art:yu+2018} stars, since the stars in the APOKASC-2 subsample represent a population subset of lower-temperature stars, as well as having lower values for \teff in the APOKASC-2 catalogue itself. However, the fact that APOKASC-2 stars represent a lower-temperature population only accounts for a shift in a measured median absolute magnitude of $\sim0.028\ \rm mag$ in $K$ and $\sim0.12\ \rm mag$ in $G$. The use of APOKASC-2 temperatures for the subset shifts the absolute magnitudes even fainter, by another $\sim 0.028\ \rm mag$ and $\sim 0.07\ \rm mag$ in $K$ and $G$, respectively. \up{At the precision afforded to us by hierarchical models, these shifts caused by the choice of temperatures become statistically significant.}

Due to the nature of the $K$ band minimizing the effects of metallicity on the RC spread, there is an extensive literature on the value of \murc in $K$. \new{It was found} by \cite{art:alves2000} to be $-1.62 \pm 0.03$ \citep[with a consistent measurement by][]{art:udalski2000}, but later placed at $-1.54 \pm 0.04$ by \cite{art:groenewegen2008}. A recent review by \cite{art:girardi2016} found a median literature value of $-1.59 \pm 0.04\ \rm mag$, which was applied by \cite{art:davies+2017} to calibrate TGAS parallaxes. New work by \cite{art:chen+2017} has used RC stars similarly identified using asteroseismology to find $-1.626 \pm 0.057\ \rm mag$, and the precursor to our hierarchical Bayesian approach, \citetalias{art:hawkins+2017}, used TGAS parallaxes to find $-1.61 \pm 0.01\ \rm mag$. Using the same method, \citetalias{art:hawkins+2017} reported an absolute magnitude of $0.44 \pm 0.01\ \rm mag$ in the \gaia $G$ band.

Our RC magnitudes for both the $K$ and \gaia $G$ bands are much closer to those reported in literature when we used APOKASC-2 stars and temperatures alone. For the $K$ band, we found values within 1$\sigma$ of \cite{art:chen+2017} for $\Delta\teff \leq 20\ K$ when using corrections to the scaling relations, although our results are otherwise incompatible with the literature for $K$. In the $G$ band, however, we found values for \murc compatible with \citetalias{art:hawkins+2017} when using APOKASC-2 stars for $\Delta\teff$ of $0$ or $+10\ K$ both with and without corrections to the scaling relations. \up{The disagreement found only in the $K$ band} could be due to our choice of bolometric corrections or corrections to the scaling relations, or due to \citetalias{art:hawkins+2017}'s choice of extinction coefficient, which is twice as large as the coefficient we use in our \gaia models, and would bias the absolute magnitudes of their stars towards brighter values. Alternatively, it could be due to \citetalias{art:hawkins+2017} not accounting for known spatial correlations in parallax \citep{art:lindegren+2016, art:zinn+2017} or possible parallax zero-point offsets \citep{art:brown2018}.

In Tables \ref{tab:parallax_k} \& \ref{tab:parallax_g}, we attempt to recreate the \citetalias{art:hawkins+2017} work, albeit including parallax covariances, and find values for \murc that are compatible with a temperature offset of $\Delta\teff < -50\ K$ for both photometric bands. \up{Finally, allowing \oozp to vary as a free parameter with loose prior constraints finds $\murc = -1.634 \pm 0.018\ \rm mag$ in the $K$ band and $0.546 \pm 0.016\ \rm mag$ in the $G$ band. These values imply that a shift to the temperature scales of $-50\ K$ or more is appropriate when using temperatures for seismology of the Red Clump. }

\subsection{Spread of the Red Clump}\label{ssec:spread}

In principle, the spread of the RC, like its luminosity, is a property of a RC population and depends on the mass and metallicity of the sample \citep{art:girardi2016,art:salaris+girardi2002}. Our hierarchical approach allows us to study the `true' spread of the RC, by evaluating the uncertainties on individual measures of absolute magnitude.

As seen for the $K$ band in Tables \ref{tab:yu_k} \& \ref{tab:apo_k}, the spread of the RC is consistent within $1\sigma$ for all perturbations of temperature, corrections to the scaling relations, and between both the \cite{art:yu+2018} and APOKASC-2 temperatures. This indicates that \sigrc is only weakly dependent on the choice of temperature scale, and that any effects of the APOKASC-2 sample only representing a small subset in metallicity are minimal for the $K$ band.



The spread of the RC due to mass and metallicity is minimised in the 2MASS $K$ band \citep{art:salaris+girardi2002}, which would lead us to expect a broader spread of the RC in the \gaia $G$ band. We see this effect in Tables \ref{tab:yu_g} \& \ref{tab:apo_g}, where the reported spreads are $\sim$ 4 to 6 times larger in magnitude. Surprisingly, we do not see the same consistency for the values of \sigrc for the $G$ band, but instead find that the inferred value of \sigrc varies inversely with temperature beyond $1\sigma$ \new{from $-50\ K$ to $50\ K$}. This trend of \sigrc with $\Delta\teff$ is likely to be an effect of the bolometric correction, as we do not see a compatible trend in $K$. \up{It should also be noted that we would expect extinction to play a larger role in the $G$ band, possibly contributing to this effect.}

For the \gaia $G$ band we also see that the value for \sigrc is lower for the APOKASC-2 sample than for the full \citetalias{art:yu+2018} sample. This reduction is liklely because the APOKASC-2 sample draws temperatures from a uniform spectroscopy source (and thus temperature scale) whereas the \citetalias{art:yu+2018} temperatures come from a variety of sources, broadening the distribution of RC stars.


The similar hierarchical approach taken by \citetalias{art:hawkins+2017} found a spread of $0.17 \pm 0.02\ \rm mag$ in $K$ and $0.20 \pm 0.02\ \rm mag$ in $G$ using TGAS parallaxes. \new{The agreement within $1\sigma$ for the $G$ band for the \citetalias{art:yu+2018} sample agrees with the inferred APOKASC-2 spread being an underestimate}. The estimates found in our work for \sigrc in $K$ are an order of magnitude lower. This is probably due to our sample size (increased from \citetalias{art:hawkins+2017} by a factor of 5) and asteroseismology providing more precise measurements for these stars than TGAS \citep{art:davies+2017}, allowing the hierarchical method to more closely constrain the true underlying spread.

Tables \ref{tab:parallax_k} \& \ref{tab:parallax_g} show the results of our attempt to recreate the \citetalias{art:hawkins+2017} work, accounting for parallax covariances and including a parallax zero-point offset. \nnew{Using \gaia parallaxes, we found a \sigrc in $K$ that is larger than our value from seismology. The results presented in Tables \ref{tab:gaia_yu_k} \& \ref{tab:gaia_apo_k}, where the the seismic \sigrc in $K$ has been applied as a prior on the \gaia model, show an inlier fraction $Q$ that is lower than we would expect for this sample. This implies that \gaia DR2 is underestimating the uncertainties for stars considered `outliers', and not including them in the inlier population.}

For the $G$ band, we found a value for \sigrc in agreement with our seismic value using APOKASC-2 temperatures. In this instance, as opposed to the results shown in Table \ref{tab:gaia_apo_g} at similar \sigrc, we find an inlier fraction $Q$ in the expected range. This is probably due to the simultaneous inference of a more appropriate value for \murc, which is closer to values established in literature \citepalias{art:hawkins+2017}. For this reason, the spreads reported in Tables \ref{tab:apo_g} \& \ref{tab:parallax_g} are our best estimates for the `true' spread of the RC in the $G$ band.

With our measurement of $\sigrc = 0.03\ \rm mag$ in the $K$ band, we can use standard error propagation through equation \ref{eq:magnitudeequation} (setting extinction to zero) to find that this spread yields a precision in distance of $\sim 1\%$ for our sample, subject to mass and metallicity. This is a factor of 5 improvement from the precision reported by \citetalias{art:hawkins+2017}. When using $\sigrc = 0.14\ \rm mag$ for the $G$ band we find a distance precision of $\sim 6\%$, in line with the findings by \citetalias{art:hawkins+2017}. 

\subsection{The \gaia parallax zero-point offset}\label{ssec:zerpointoffset}
The \gaia DR2 parallax zero-point offset, while small, can still have an effect on results, and is widely applied in studies using DR2 \citep{art:luri+2018, art:bailer-jones+2018}, with potentially far-reaching consequences \citep{art:shanks+2018}. The offset has been estimated through calibration with eclipsing binaries \citep{art:stassun+torres2018}, Cepheids \citep{art:riess+2018}, asteroseismology \citep{art:zinn+2018,art:sahlholdt+silvaaguirre2018}, kinematics \citep{art:schonrich+19} and quasars \citep{art:lindegren+2018}.

In Tables \ref{tab:gaia_yu_k}, \ref{tab:gaia_apo_k}, \ref{tab:gaia_yu_g} and \ref{tab:gaia_apo_g} we present our inferred model parameters given our values for \murc and \sigrc found through asteroseismology at different temperature shifts $\Delta\teff$, effectively `calibrating' \gaia DR2 to see what offset recovers a given set of RC parameters. 

\nnew{Figure \ref{fig:gaia_posteriors} shows the posterior distributions for \oozp given our seismic priors from different temperature shifts, where there is a clear trend of \oozp with seismic \murc, and thus with temperature. This trend was also found in recent results by \cite{art:khan+2019}, where a comparison of \gaia parallaxes and seismic distances obtained through the seismic scaling relations found that a temperature shift of $100\ K$ caused a shift in \oozp of $10-15\ \muas$ for RC stars, although it should be noted that they found this effect largely reduced when using grid modelling techniques \citep{art:rodrigues+2017}.}

It is also apparent in Figure \ref{fig:gaia_posteriors} that the uncertainty on \oozp is significant, and consistent for all model conditions, due to the parallax covariances presenting a systematic lower limit on parallax uncertainties for this sample. Given a \murc in $K$ closer to literature values, with the run corresponding to APOKASC-2 temperatures using $\Delta\teff = -50\ K$, we found a \oozp within $1\sigma$ of the uncertainties on all literature values for \oozp in the \kepler field discussed in this work. This is both an encouraging sign of a consistent \oozp in the \kepler field, and further indication that seismology would be improved by reducing the temperature scale. For the \gaia $G$ band, the run closest to the existing literature ($\Delta\teff = 0$) is consistent with all values for \oozp besides \cite{art:stassun+torres2018}.

Given a selection of values for \oozp reported in the literature, we applied informative priors on \oozp in our \gaia model, and allowed \murc and \sigrc to explore the parameter space freely. The results of this are shown in Tables \ref{tab:parallax_k} \& \ref{tab:parallax_g}, for the $K$ and $G$ bands respectively. The credible intervals for \murc are shown in Figure \ref{fig:parallax_values}. For both bands, we found that the choice of \oozp from the literature had no impact beyond $1\sigma$ on either of the RC properties for any values used. When using a tightly constrained \oozp of zero (in an attempt to recreate \citetalias{art:hawkins+2017}) we found the largest overall change. It is also interesting to note that for a prior corresponding to the \cite{art:stassun+torres2018} value, the inferred value for \oozp is reduced to lie closer to those found in other works for the \kepler field.

\new{Finally, running the \gaia model with uninformative priors on both \oozp and the RC parameters produced a parallax zero-point offset of $(-38\pm13)\ \muas$ in $K$ and $(-42\pm13)\ \muas$ in $G$ for the \kepler field. These values are consistent with one another and with the existing literature, and also agree with recent results by \cite{art:khan+2019} for RC stars in APOKASC-2.} Given the uncertainties on the inferred values of \oozp, we see a fundamental uncertainty limit on \gaia parallaxes of $\sim 13\ \mu \rm as$ as a result of spatial covariances in the parallaxes. \new{Encouragingly, this implies that for our RC sample in the \kepler field, the choice of parallax-zero point offset \nnew{does not dramatically impact} the inferred luminosities, given a proper treatment of the spatial parallax covariances. However, this may not generalize to populations more sparsely sampled in space, and in other magnitude ranges, given the known relation between the parallax zero-point offset, $G$ band magnitude and colour \citep{art:zinn+2018,art:lindegren+2018}.}


\subsection{Corrections to the seismic scaling relations}\label{ssec:scalingrelations}

In Section \ref{ssec:zerpointoffset}, we have compared results with and without corrections to the \dnu seismic scaling relation, \fdnu, derived from \cite{art:sharma+stello2016}. It is known that stellar models do not not accurately reproduce the \dnu of the Sun (off by about $1\%$), due to the so-called surface effect \citep{art:christensen-dalsgaard+1988, art:white+2011}. Corrections to the scaling relation \fdnu derived without accounting for the surface effect \citep[such as ][]{art:sharma+stello2016} can produce radii that differ on the order of $\sim 2\%$ from methods that do \citep[such as][]{art:rodrigues+2017}. As a check, we considered the impact that this may have on our inferred values for the RC magnitude.

To compare the calculated RC populations in the $K$ and \gaia $G$ bands, we obtained radii using \fdnu obtained through \cite{art:sharma+stello2016}. We then used bolometric corrections for no temperature offset to calculate the absolute magnitudes using both those radii and those same radii reduced by both $1.6\%$ and $2.4\%$. We found that a reduction on radius in the range $(2 \pm 0.4)\%$ resulted in a global shift toward brighter bolometric magnitudes by $44^{+9}_{-8}\ \rm mmag$.

In Tables \ref{tab:apo_k} \& \ref{tab:apo_g} we report the absolute magnitude of the RC (for no temperature offset) in the APOKASC-2, \fdnu-corrected, sample of $-1.69\ \rm mag$ in $K$ and $0.45\ \rm mag$ in $G$. A shift of $0.04\ \rm mag$ applied to both bands is enough to reconcile our seismic results with those obtained through \gaia for both the $K$ and $G$ bands, as well as those from the literature. Note however that this is not the case when applied to the \citetalias{art:yu+2018} sample (see Tables \ref{tab:yu_k} \& \ref{tab:yu_g}), where this shift applied in both bands would not be enough to reconcile the seismic results for the absolute magnitude of the RC with any measures both in this work or in the literature.

\subsection{Calibrating \gaia and asteroseismology}
Our initial aim with this work was to calibrate the \gaia parallax zero-point offset, \oozp, using asteroseismology. Given the large change in the absolute magnitude of the RC, \murc, with relatively small changes in temperature for our large RC population, and consequently the shift in inferred \oozp given these values for \murc, it proved difficult to definitively calibrate \gaia parallaxes using seismology.

The reverse however, seems more possible. We found that the various parallax offsets reported in the literature, when used as informative priors on our \gaia model, all resulted in similar values for \murc in both the 2MASS $K$ and \gaia $G$ bands (as shown in Tables \ref{tab:parallax_k} \& \ref{tab:parallax_g}), and inferred values for \oozp that lie closer together for those literature values with large uncertainties \citep{art:stassun+torres2018,art:riess+2018,art:sahlholdt+silvaaguirre2018}. Imposing a prior for \oozp to lie close to zero showed a departure beyond $1\sigma$ from the \murc values found otherwise, indicating that \oozp does have a measurable effect on the inferred RC luminosity. Finally, applying no strongly informative priors on the RC parameters nor \oozp led to inferred values of \murc and \oozp being consistent with values in the literature, albeit with a large uncertainty of $\sim 13\ \mu \rm as$ on the parallax zero-point offset, implying a fundamental limit on the uncertainty on this offset given the spatial parallax covariances.

Given that the choice of parallax zero-point offset did not dramatically affect the inferred luminosity of the clump \new{(see Tables \ref{tab:parallax_k} \& \ref{tab:parallax_g} and Figure \ref{fig:parallax_values})}, we can reasonably use any value of \oozp reported in the literature, including from this work, to attempt a calibration of seismology. Given the results for our runs on \gaia data with RC parameters constrained by seismology (Tables \ref{tab:gaia_apo_k} \& \ref{tab:gaia_apo_g}), we expect that $\murc = -1.634\ \rm mag$ in $K$ and in $0.546\ \rm mag$ in $G$ would be roughly consistent with a temperature offset $\Delta\teff$ between $\sim -100\ K$ and $\sim - 70\ K$ for temperatures in the APOKASC-2 catalogue \citep[(which, as has been noted, are already lower than those reported by][for the same stars]{art:mathur+2017}. An offset of this size would fall within known systematic uncertainties on temperatures inferred from seismology \citep{art:slumstrup+2018}. However, it should be noted that this shift in temperature scale is degenerate with the scaling relations underestimating radii by $\sim 2\%$ compared to our estimates for radius using corrections by \cite{art:sharma+stello2016}, as discussed in section \ref{ssec:scalingrelations}.

In order to confirm these proposed shifts to temperature, we reran our asteroseismic model on our APOKASC-2 subsample for a range of temperature shifts extended down to $-110 K$ for both the $K$ and \gaia $G$ bands, with RC-corrected scaling relations. We found that when considering the $K$ band, our calibration value for \murc from \gaia corresponds to within $1\sigma$ with a temperature shift of between $-110$ and $-70\ K$. When considering the $G$ band, the \gaia \murc corresponds to within $1\sigma$ for a shift between $-70$ and $-50\ K$. Given that any calibrated correction to the temperature scale should be applied globally to the full APOKASC-2 subsample, we find that a temperature shift of $-70\ K$ to the temperatures of our RC subsample of APOKASC-2 would produce seismic absolute magnitudes of the clump consistent with those found using \gaia DR2.

\nnew{We only ran this test for the APOKASC-2 subsample, for which temperatures were all drawn from a uniform spectroscopic source. Since the temperatures for the full \citetalias{art:yu+2018} are not, claims about changes to temperature scales for this sample would be inappropriate.}

The ability to make this inference reliably rests on our hierarchical treatment, as initially set out by \citetalias{art:hawkins+2017}, and treatment of the spatial correlations in parallax reported by \cite{art:lindegren+2018}. As we improve our understanding of these correlations, our inferences using this and similar hierarchical models will improve. Similarly, it is known that population effects in age, metallicity and temperature, among others, have an effect on the inferred luminosity of the RC \cite{art:girardi2016}. Our hierarchical model, can be further improved by accounting for these effects, as well as including parameters that check for consistent colours, as suggested by \citetalias{art:hawkins+2017}. As these hierarchical models improve in future work, so will our understanding of the RC, and our ability to calibrate asteroseismology.

\section{Conclusions} \label{sec:conclusions}
Using two hierarchical models, based on the work by \citetalias{art:hawkins+2017}, we inferred the spread and position in absolute magnitude of a sample of \nstars Red Clump (RC) stars in the 2MASS $K$ and \gaia $G$ bands. We first did this using absolute magnitudes obtained through a completely distance-independent asteroseismic method, and probed systematics in asteroseismology by varying the temperatures of the sample, applying corrections to the scaling relations, and running our model on a subsample of stars with separate spectroscopic temperatures reported in APOKASC-2 \citep{art:pinsonneault+2018}. We then applied the results from seismology as strongly informative priors on the position and spread of the clump for our second hierarchical model. We applied this to \gaia DR2 data in order to see how the parallax zero-point varied, taking into account spatial correlations of parallaxes reported by \cite{art:lindegren+2018}. We then applied strongly informative priors on the parallax zero-point in our \gaia model and allowed the RC parameters to roam more freely, to study the impact of published values for the zero-point offset on the RC. Finally, we performed a run of the \gaia model with no strongly informative priors on any parameters.

We leave the reader with the following conclusions:

\begin{enumerate}
\item By applying the \citetalias{art:hawkins+2017} hierarchical model, with improvements to account for spatial correlations of parallaxes and maginalize over the parallax zero-point offset (\oozp), \up{we find a mean value for \oozp in the \kepler field to be $-41 \pm 10\ \mu \rm as$ for our sample}, consistent with all existing measures of \oozp in the \kepler field. This offset results in a Red Clump magnitude of $-1.634 \pm 0.018$ in $K$ and $0.546 \pm 0.016$ in $G$ for our sample.

\item Applying a hierarchical model to our sample of absolute magnitudes obtained from asteroseismology, we find a spread of the RC in the 2MASS $K$ band of $\sim 0.03\ \rm mag$ independent of our changes made to the sample, an order of magnitude lower than the value reported previously using \gaia TGAS parallaxes in \citetalias{art:hawkins+2017}. This extremely small spread highlights the power of seismology and the potential of the RC in the $K$ band as a standard candle. In the \gaia $G$ band we find a spread of $\sim 0.13\ \rm mag$ using APOKASC-2 temperatures, which is consistent with results found using \gaia DR2 parallaxes.

\item We find that a small global change in temperature ($\sim 10\ - 20\ K$) can affect the inferred absolute magnitude of the RC from seismology by more than $1\sigma$, and is degenerate with the application of a correction \fdnu to the seismic scaling relations.

\nnew{\item We find values for the absolute magnitude of the RC from seismology to agree within $1\sigma$ with those inferred from \gaia DR2 parallaxes in both the $K$ and $G$ bands, only if a global temperature shift of $\sim -70\ K$ is applied to our RC subsample of APOKASC-2 stars. This shift is within expected systematic uncertainties on spectroscopic techniques. These differences are also degenerate with a shift in seismic radius of $2\%$, which is within the uncertainty imposed by choice of corrections to the scaling relations.}

\item A hierarchical Bayesian mixture model for a population of RC stars, as first set out by \citetalias{art:hawkins+2017}, continues to be an excellent tool for working with \gaia DR2 parallaxes, with the new additions of a parallax zero-point offset as a parameter and spatial correlations between parallaxes. Further additions will undoubtedly improve our inferences on RC stars, and with it, our ability to calibrate asteroseismology and \gaia.
\end{enumerate}

\section*{Acknowledgements}
The authors would like to thank the anonymous reviewer for their helpful comments, which contributed to the quality of this manuscript. They would also like to thank Beno\^it Mosser, Marc Pinsonneault, Joel Zinn and James Kuszlewicz for the helpful discussions. 
OJH, GRD, YPE and WJC acknowledge the support of the UK Science and Technology Facilities Council (STFC).
AM acknowledges support from the ERC Consolidator Grant funding scheme (project ASTEROCHRONOMETRY, G.A. n. 772293).
AGAB acknowledges financial support from the Netherlands Research School for Astronomy (NOVA).
KH is partially supported by a Research Corporation TDA Grant. 
RAG acknowledges the support from the PLATO CNES grant.
This work has made use of data from the European Space Agency (ESA) mission {\it Gaia} (\url{https://www.cosmos.esa.int/gaia}), processed by the {\it Gaia}.
Data Processing and Analysis Consortium (DPAC, \url{https://www.cosmos.esa.int/web/gaia/dpac/consortium}). Funding for the DPAC has been provided by national institutions, in particular the institutions participating in the {\it Gaia} Multilateral Agreement. This publication makes use of data products from the Two Micron All Sky Survey, which is a joint project of the University of Massachusetts and the Infrared Processing and Analysis Center/California Institute of Technology, funded by the National Aeronautics and Space Administration and the National Science Foundation.

\textsc{Software:} This work has made use of the following software not cited in the text: \texttt{astropy} \citep{art:astropycollaboration+2013, art:astropycollaboration+2018}, \texttt{corner} \citep{sw:corner}, \texttt{daft} \citep{sw:daft}, \texttt{PyStan} \citep{sw:pystan}, \texttt{Stan} \citep{art:carpenter+2017}, \texttt{scikit-learn} \citep{sw:scikit}, \texttt{iPython} \citep{sw:ipython}, \texttt{Jupyter Notebooks} \citep{sw:jupyter}, \texttt{numpy} \citep{sw:numpy}, \texttt{pandas} \citep{sw:pandas}, \texttt{matplotlib} \citep{sw:matplotlib} and \texttt{seaborn} \citep{sw:seaborn}.



\bibliographystyle{mnras}
\bibliography{halletal18_na} 

\bsp	
\label{lastpage}
\end{document}